\documentclass[10pt]{article}
 
\usepackage[T1]{fontenc}
\usepackage[latin1]{inputenc} 
\usepackage{amssymb}
\usepackage{graphicx}
\usepackage{booktabs}           %better tables
\usepackage{url}
\usepackage{ae,aecompl}
\usepackage[round]{natbib}
\usepackage{amsmath}
\usepackage[body={6in,8.4in},top=1.4in,left=1.3in]{geometry}
\usepackage{fancyheadings}
\newcommand{\bra}{\left\langle}
\newcommand{\ket}{\right\rangle}
\newcommand{\veca}{\mathbf{a}}
\newcommand{\vecw}{\mathbf{w}}
\newcommand{\vecs}{\mathbf{s}}
\newcommand{\vecu}{\mathbf{u}}
\newcommand{\vecv}{\mathbf{v}}
\newcommand{\vecx}{\mathbf{x}}

\newcommand{\matA}{\mathbf{A}}

\newcommand{\matW}{\mathbf{W}}

\newcommand{\pP}{\mathcal{P}}

\newcommand{\wij}{w_{ij}}

\newcommand{\sign}{\mathrm{sign}}

\newcommand{\imagi}{\mathfrak{i}}

\begin{document}

\title{\textsf{Complex Independent Component Analysis of Frequency-Domain Electroencephalographic Data}}
\author{Jörn Anemüller, Terrence J.\ Sejnowski and Scott Makeig\\
  \\ \small Swartz Center for Computational Neuroscience, Institute for Neural Computation\\
  \small University of California San Diego, 9500 Gilman Dr., Dept.\ 0961, La Jolla, CA 92093-0961, USA\\
  \small and\\
  \small Computational Neurobiology Laboratory, The Salk Institute for Biological Studies\\
  \small 10010 N. Torrey Pines Rd., La Jolla, CA 92037, USA
}

\date{\small\today}

\maketitle

\thispagestyle{empty}

\begin{center}
  \small
  Published in: \emph{Neural Networks}, 16:1311--1323, 2003.\vspace{1.2em}
\end{center}

\begin{abstract}
  Independent component analysis (ICA) has proven useful for modeling brain and electroencephalographic (EEG) data.
  Here, we present a new, generalized method to better capture the dynamics of brain signals than previous ICA algorithms.
  We regard EEG sources as eliciting spatio-temporal activity patterns, corresponding to, e.g., trajectories of activation propagating across cortex.
  This leads to a model of convolutive signal superposition, in contrast with the commonly used instantaneous mixing model.
  In the frequency-domain, convolutive mixing is equivalent to multiplicative mixing of complex signal sources within distinct spectral bands.
  We decompose the recorded spectral-domain signals into independent components by a complex infomax ICA algorithm.
  First results from a visual attention EEG experiment exhibit 
  (1) sources of spatio-temporal dynamics in the data, 
  (2) links to subject behavior,
  (3) sources with a limited spectral extent, and
  (4) a higher degree of independence compared to sources derived by standard ICA.
\end{abstract}
\bigskip

\tableofcontents
\bigskip

\noindent {\bf Keywords:}
complex independent component analysis (ICA), frequency-domain, convolutive mixing, biomedical signal analysis, electroencephalogram (EEG), event-related potential (ERP), visual selective attention
\bigskip

\noindent {\bf Acknowledgments:}
J.~A.~was supported by the German Academic Exchange Service (DAAD) and the German Research Council (DFG).
We acknowledge support from the Swartz Foundation.

\lhead{\it Anemüller, Sejnowski, Makeig: Complex ICA of Frequency-Domain EEG Data}

\newpage

\section{Introduction}
\label{sec:introduction}

Independent component analysis (ICA) is effective in analyzing brain signals and in particular electroencephalographic (EEG) data \citep[e.g.,][]{makeig96:_indep_compon_analy_elect_data,makeig02:_dynam_brain_sourc_visual_evoked_respon}, and ICA continues to be useful for building new models of experimental data.
However, ICA algorithms presently applied to brain data rely on several idealized assumptions about the underlying processes that may not be fully applicable.
Although the results so far obtained with ICA are significant and justify its continued use, it is nevertheless desirable to advance the ICA methodology by allowing more realistic modeling of EEG dynamics.

One principal limitation imposed on ICA algorithms is the mixing process by which the source signals are assumed to be superimposed to form the measured sensor signals.
Presently, ICA analysis of brain data is carried out assuming a linear and instantaneous mixing process that can be expressed mathematically as multiplication by a single mixing matrix.
The physics of electromagnetic wave propagation support instantaneous summation at the electrode sensors since capacitive effects within the head are generally regarded as negligible at EEG frequencies of interest \citep{Lagerlund_EEG_1999}.

In the standard ICA model the component signal sources are thought of as neural activity occurring in a perfectly synchronized manner within spatially fixed cortical domains.
This assumption might be too strong, as it does not take into account the possible spatio-temporal dynamics of the underlying neural processes, e.g., propagation of neuronal activity, traveling wave patterns of activity, or synchronization between different brain areas with a non-zero phase lag \citep{Freeman_MassActNervSys_1975,LopezS_chapter_1978,ArieliSGA_Science_1996,stein00:_top}.
One way to allow the effective sources to exhibit more complex dynamics is to assume a convolutive mixing model.
In a convolutive mixing process, a single impulse-like activation of an EEG component may elicit a sequence of potential maps with varying spatial topography; such a model may thereby allow for patterns of spatial propagation of EEG activity.
Separation of convolutively mixed sources into independent EEG components is not feasible under the instantaneous mixing assumption, since the temporal autocorrelation of the EEG results in statistical dependencies between the time-courses of consecutive potential maps.
At best, instantaneous ICA may separate moving sources into separate stationary components with overlapping `frames' of activation \citep{MakeigJGS_IntHumBrainSci_2000}.

A fundamentally different phenomenon, also neglected by the standard ICA model, is the spectral quality of EEG signals.
EEG researchers have long been familiar with the fact that EEG activity has distinctive characteristics in different frequency bands (conventionally delta, theta, alpha, beta, and gamma) which may be associated with different physiological processes (\citet{Berger_ArchPsychNerv_1929}, and also \citet{MakeigI_EEGClinNeuPhys_1993}).
It may therefore be more appropriate to allow for the existence of different functionally independent sources in different frequency bands by modeling the source superposition with a different mixing matrix for each frequency band.

To overcome both shortcomings, we approach convolutive independent component analysis of EEG signals through complex ICA applied to different spectral bands.
Convolutive mixing in the time-domain is equivalent to multiplicative mixing in the frequency-domain with generally distinct complex-valued mixing coefficients in different frequency bands.
Therefore, by moving to the frequency-domain, both spatio-temporal source dynamics and frequency-specific source processes may be modeled.
Solutions obtained under the standard ICA model with instantaneous mixing in the time-domain form a subset within the solution space of complex frequency-domain ICA, corresponding to signal superposition with the same real-valued mixing matrix in all frequency bands.

\begin{figure*}[tbp]
  \centering
  \includegraphics[width=0.8\linewidth,keepaspectratio]{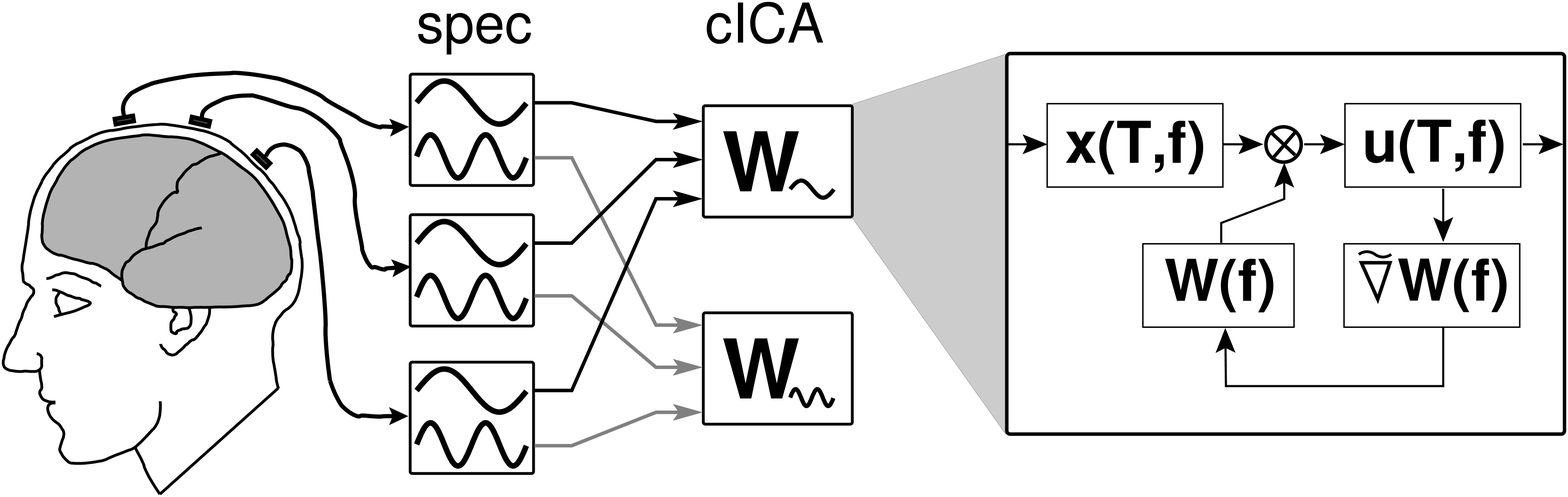}
  \caption{Schematic representation of the processing stages of the complex spectral-domain ICA algorithm. Left (`spec'): the recorded electrode signals are decomposed into different spectral bands. Center (`cICA'): Complex ICA decomposition is performed within each spectral band. Right: Iteration steps performed by complex ICA for estimation of each separating matrix $\matW(f)$.}
  \label{fig:1}
\end{figure*}

The method consists of two processing stages (cf.\ Fig.~\ref{fig:1}).
First, the measured EEG signals are decomposed into different spectral bands by short-time Fourier transformation or wavelet transformation, yielding a complex-valued spectro-temporal representation for each electrode signal.
Then, a separate independent component analysis is performed on the complex frequency-domain data within each spectral band, producing, for each band, a set of complex independent component activation time-courses and corresponding complex scalp maps.
We also investigate the case of a constrained complex ICA algorithm where the independent component (IC) activations remain complex, but the IC scalp maps are required to be real-valued.

Within each spectral band, the proposed algorithms find a number of maximally independent components equal to the number of employed data channels.
Hence, across frequencies the method has the potential of identifying more independent processes than the number of electrodes.
But since EEG processes may not be narrow-band, but may exhibit dynamics within multiple contiguous or disconnected bands, independent components at different frequencies might also originate from the same spatial EEG generator sources.
This could, e.g., be the case for mu-rhythm activity \citep{NiedermeyerL_EEG_1999} which is characterized by concurrent activity near 10~Hz and 20~Hz.
We present methods for evaluating the similarity of independent components at different frequencies and for grouping together those components arising from single physiological processes.

Convolutive mixing models have been used for blind source separation in other domains.
For example, in the case of speech signals the physics of wave propagation in air directly leads to a convolutive mixing process \citep[e.g.,][]{Anemuller_PhD_2001}.
However, speech signals are generally modeled as wide-band sources, emitting energy essentially over the entire spectral range of interest.
The same assumption cannot be made for brain signals, so that corresponding convolutive ICA algorithms cannot be applied directly to the problem at hand.
On the other hand, narrow-band sources are also encountered in telecommunications applications, leading to ICA algorithms similar to the one presented here \citep{torkkola98:_blind}.
However, a strict narrow-band assumption may not be completely justified for brain signal sources, as mentioned above.
The methods presented in this paper appear to be sufficiently flexible to model signals in all the aforementioned scenarios.

The remainder of the paper is organized as follows:
In sections~\ref{sec:spectr-decomp} and~\ref{sec:freq-band-spec}, we define the spectral decomposition and mixing model.
We derive a complex variant of the infomax ICA algorithm \citep{Bell95a} in section \ref{sec:complex-valued-ica} from the maximum-likelihood principle, and discuss a variant constrained to real scalp maps in section \ref{sec:compl-ica-constr}.
Visualization of complex activations and maps is discussed in \ref{sec:visu-comp-maps}.
Section \ref{sec:assess-qual-separ} defines second and fourth order measures for assessing the quality of the separation.
In section \ref{sec:align-source-comp} we present methods for measuring similarities between independent components in distinct spectral bands.
Section~\ref{sec:time-domain-ica} presents methods for comparing real time-domain and complex frequency-domain ICA results.
Finally, we apply these methods to data from a visual attention task EEG experiment in section \ref{sec:results}.

\section{Methods}
\label{sec:methods}

\subsection{Spectral decomposition}
\label{sec:spectr-decomp}

Consider measured signals $x_i(t)$, where $i=1,\ldots,M$ denotes electrodes.
Their spectral time-frequency representations are computed as
\begin{equation}
  \label{eq:1}
  x_i(T,f) = \sum_{\tau} x_i(T+\tau) b_f(\tau),
\end{equation}
where $f$ denotes center frequency, and $b_f(\tau)$ the basis function which extracts the spectral band $f$ from the time-domain signal.
The basis function is centered at time $T$.
Hence, data of size 
$\left[ \mathrm{channels}\ i \,\times\, \mathrm{times}\ t \right]$ is transformed into data of size 
$\left[ \mathrm{channels}\ i \,\times\, \mathrm{times}\ T \,\times\right.$ $\left.\mathrm{frequencies}\ f \right]$.

In this paper, we consider the decomposition by means of the short-time Fourier transformation, such that $b_f(\tau)$ is given by
\begin{equation}
  \label{eq:2}
  b_f(\tau) = h(\tau) \, e^{-\imagi 2\pi f \tau/2K},
\end{equation}
$h(\tau)$ being a windowing function (e.g., a hanning window) with finite support in the interval $\tau=-K,\ldots,K-1$, and $2K$ denoting the window length.
Correspondingly, the frequency index acquires values $f=0,\ldots,K$.
Since the product of time- and frequency-resolution is bounded from below by $0.5$, the chosen windowing function and window length give limited frequency-domain resolution.
Hence, variability across frequencies is limited and results should be interpreted accordingly.

\subsection{Mixing model}
\label{sec:freq-band-spec}

For each frequency band $f$ the signals $\vecx(T,f) = [x_1(T,f),$ $\ldots, x_M(T,f)]^T$ are assumed to be generated from independent sources $\vecs(T,f) = [s_1(T,f),\ldots,s_N(T,f)]^T$ by multiplication with a frequency-specific mixing matrix $\matA(f)$,
\begin{equation}
  \label{eq:3}
  \vecx(T,f) = \matA(f) \, \vecs(T,f),
\end{equation}
with $\mathrm{rank}(\matA(f))=N$.
Noise is assumed to be negligible.
We restrict the presentation to square-mixing, $M=N$, though our methods are also applicable to the case $M>N$.
The estimates $\vecu(T,f)$ of the sources are obtained from the sensor signals by multiplication with frequency-specific separating matrices $\matW(f)$,
\begin{equation}
  \label{eq:4}
  \vecu(T,f) = \matW(f) \, \vecx(T,f).
\end{equation}

\subsection{Complex ICA}
\label{sec:complex-valued-ica}

\begin{figure}[t]
  \centering
  \includegraphics[width=0.45\linewidth,keepaspectratio]{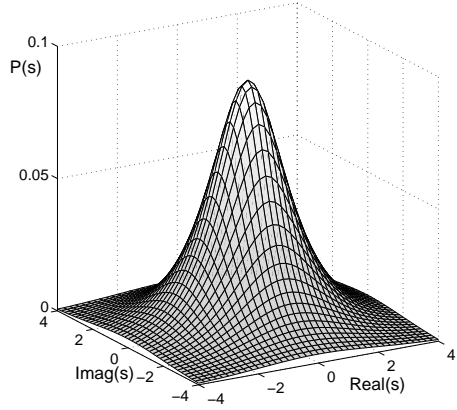}
  \caption{The circular symmetric super-Gaussian probability density function $\pP(s)$ of the complex sources $s$.}
  \label{fig:2dimpdf}
\end{figure}

\begin{figure}[t]
  \centering
  \includegraphics[width=0.45\linewidth,keepaspectratio]{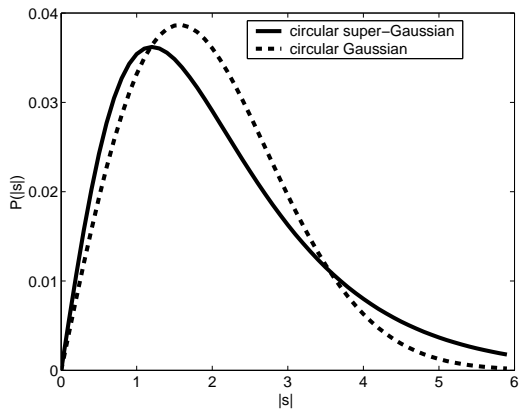}
  \caption{The distribution $\pP_{|s|}(|s|)$ of super-Gaussian source magnitude (solid) versus the distribution of the magnitude of a two-dimensional Gaussian process with the same variance (dashed). 
    The latter is the well-known Rayleigh distribution. 
    The super-Gaussian source distribution is characterized by its stronger peak at small magnitudes and its longer (high-magnitude) tails.}
  \label{fig:nongaussian}
\end{figure}

To derive the complex infomax ICA algorithm, we model the sources $s_i(T,f)$ as complex random variables with a circular symmetric, non-Gaussian probability density function $\pP_s(s)$.
Since the phase $\arg(s_i(T,f))$ depends only on the relative position of the window centers $T$ with respect to the time-domain signal $s_i(t)$, the property of circular symmetry of the distribution $\pP_s(s)$ is a direct result of the window-centers being chosen independently of the signal.
Hence, $\pP_s(s)$ depends only on the magnitude $|s|$ of $s$ and can be written as
\begin{equation}
  \label{eq:8}
  \pP_s(s)=g(|s|)
\end{equation}
with the function $g(\cdot)\, : \, \mathbb{R}\rightarrow\mathbb{R}$ being a real-valued function of its real argument.
Our investigation of the statistics of frequency-domain EEG in section \ref{sec:kurtosis} demonstrates the data's positive kurtosis.
Therefore, we choose $\pP_s(s)$ as a super-Gaussian distribution.
The assumed two-dimensional distribution $\pP_s(s)$ over the complex plane is illustrated in Fig.~\ref{fig:2dimpdf}.
Its super-Gaussian nature is best seen by plotting the corresponding distribution $\pP_{|s|}(|s|)$ of the \emph{magnitude} $|s|$ versus the corresponding distribution (a Rayleigh distribution) for a two-dimensional Gaussian distribution of the same variance, as illustrated in Fig.~\ref{fig:nongaussian}.

The separating matrix $\matW(f)$ is obtained by maximizing the log-likelihood $L(\matW(f))$ of the measured signals $\vecx(T,f)$ given $\matW(f)$, which in terms of the source distribution $\pP_s$ is
\begin{equation}
  \label{eq:18}
  L(\matW(f)) = \bra \log \pP_x (\vecx(T,f) | \matW(f)) \ket_T 
              = \log \det (\matW(f)) + \bra \log \pP_s \left( \matW(f) \, \vecx(T,f) \right) \ket_T,
\end{equation}
where $\bra \cdot \ket_T$ denotes expectation computed as the sample average over $T$.
We perform maximization by complex gradient ascent on the likelihood-surface.
The $(i,j)$-element $\delta w_{ij}(f)$ of the gradient matrix $\nabla \matW(f)$ is defined as
\begin{equation}
  \delta \wij (f)
  = 
  \left( \frac{\partial}{\partial \Re \wij(f)} + \imagi \frac{\partial}{\partial \Im \wij(f)} \right) L(\matW(f)),
\end{equation}
where $\partial/\partial \Re \wij(f)$ and $\partial/\partial \Im \wij(f)$ denote differentiation with respect to the real and imaginary parts of matrix element $\wij(f)=\left[ \matW(f) \right]_{ij}$, respectively.
This results in the gradient
\begin{equation}
  \label{eq:20}
  \mathbf{\nabla W}(f) = \left( \mathbf{I} - \bra \vecv(T,f) \, \vecu(T,f)^H \ket_T \right) \matW^{-H}(f),
\end{equation}
however, faster convergence is achieved by using the complex extension of the natural gradient \citep{Amari96a}
\begin{equation}
  \label{eq:6}
  \mathbf{\tilde{\nabla} W}(f) = 
  \mathbf{\nabla W}(f) \, \matW(f)^H \, \matW(f) = 
  \left( \mathbf{I} - \bra \vecv(T,f) \, \vecu(T,f)^H \ket_T \right) \matW(f),
\end{equation}
where $\vecv(T,f)$ is a non-linear function of the source estimates $\vecu(T,f)$:
\begin{align}
  \label{eq:7}
  \vecv(T,f) &= [v_1(T,f),\ldots,v_N(T,f)]^T,\\
  v_i(T,f)   &= \sign(u_i(T,f)) \, \frac{g^\prime(|u_i(T,f)|)}{g(|u_i(T,f)|)},\\
  \sign(z)   &= 
  \begin{cases}
    0     & \text{if $z=0$,}\\
    z/|z| & \text{if $z \neq 0$.}
  \end{cases}
\end{align}
Here, $\mathbf{I}$ denotes the identity matrix, $g^\prime(\cdot)$ is the first derivative of function $g(\cdot)$, and $H$ denotes complex conjugation and transposition.
The gradient Eq.\ \eqref{eq:6} was previously used in an algorithm for blind separation of speech signals \citep{AnemullerK_SpeCom_2003}.

For the choice
\begin{equation}
  \label{eq:9}
  \frac{g^\prime(x)}{g(x)} = \frac{1-e^{-x}}{1+e^{-x}}
\end{equation}
we obtain a complex generalization of the standard logistic infomax ICA learning rule \citep{Bell95a}.
The algorithm may be adapted to different (e.g., sub-Gaussian) source distributions by use of other appropriate non-linearities $g^\prime/g$.
In the case of purely real-valued data, the learning rule for complex data reduces to the infomax ICA learning rule for real signals.

Due to the circular symmetry of $\pP_s$, the log-likelihood $L(\matW(f))$ is invariant with respect to the multiplication of any row $\vecw_i(f)$ of $\matW(f)$ with an arbitrary unit-norm complex number $c_i(f)$, $|c_i(f)|=1$.
This parallels the sign-ambiguity of real ICA algorithms using symmetric non-linearities.
However, since the circular symmetry allows for continuous invariance transformations (in contrast to the discrete sign-flip operation), detection of convergence is hindered.
Therefore, we constrain the diagonal of $\mathbf{\tilde{\nabla} W}(f)$ by projecting it to the real line, thereby reducing the invariance to a sign-flip ambiguity.

The independent component decomposition based on Eq.\ \eqref{eq:6} is performed separately for each frequency band $f$, yielding in total $N(K+1)$ complex independent component activation time-courses $u_j(T,f)$ and $N(K+1)$ complex scalp maps $\veca_j(f)$, where $\veca_j(f)$ denotes the $j$-th column of the estimated mixing matrix $\matA(f)=\matW^{-1}(f)$.

\subsubsection{Complex ICA constrained to real scalp maps}
\label{sec:compl-ica-constr}

The complex scalp maps $\veca_j(f)$ can be interpreted in terms of amplitude- and phase-differences between different spatial positions on the scalp produced by the spatio-temporal dynamics of the underlying EEG generators.
However, it might also be of value to constrain the scalp maps to be real-valued as in standard ICA.
In this constrained model of the EEG, sources are assumed to be frequency-specific (in contrast to the wide-band source model of standard ICA), but may not elicit the spatio-temporal dynamics of the fully-complex model.
Together with a simpler interpretation, this approach has the advantage of making it possible to further separate the effects induced by wide-band versus band-limited data and by instantaneous (real) versus convolutive (complex) mixing.
\footnote{Mathematically, signal superposition by means of different real-valued mixing matrices in distinct frequency bands can also be interpreted as convolutive mixing of wide-band sources, but with symmetric filters.
However, this special case of convolutive mixing may be too restricted to fully model the possible complexity of the underlying neuronal dynamics.
Therefore, we adopt the more plausible interpretation that real-valued mixing in different frequency bands reflects band-limited processes without spatio-temporal dynamics.}

To constrain the algorithm's solution to real scalp maps, the initial estimate of $\matW(f)$ is chosen to be real (typically the identity matrix), and the gradient Eq.\ \eqref{eq:6} is projected to the real plane, resulting in the constrained gradient
\begin{equation}
  \label{eq:11}
  \mathbf{\tilde{\nabla}_\Re W}(f) = \Re \left( \mathbf{\tilde{\nabla} W}(f) \right),
\end{equation}
with $\Re$ denoting the real part.
While the corresponding scalp maps $\veca_j(f)$ are real, the separated IC activations $\vecu(T,f)$ remain complex.

Eq.\ \eqref{eq:11} differs from, e.g., applying standard infomax ICA to the real-parts of $\vecu(T,f)$ in that its underlying source model Eq.\ \eqref{eq:8} is still based on a distribution over the complex plane.
As a result, the product $\vecv(T,f)\,\vecu(T,f)^H$ in the right hand side of Eq.\ \eqref{eq:11} is evaluated using complex multiplication.
In principle, performing complex ICA to derive  real-valued component maps might be more accurate than performing real ICA on concatenated real and imaginary parts of band-limited time-frequency transformations as proposed by \citet{ZibulevskyKZP_NIPS_2002} since the circular symmetric complex distribution assumed by complex ICA should be more accurate than the assumption of mutual independence between real and imaginary parts used in the real spectral ICA decomposition method.

\subsubsection{Visualizing complex IC activations and maps}
\label{sec:visu-comp-maps}

Complex independent component activations $u_i(T,f)$ may be conveniently visualized by separately plotting their power (squared amplitude) and phase.
To simplify the visual impression of the phase data, we compensate for the effect of phase-advances locked to the carrier frequency by complex demodulation \citep[e.g.,][]{Bloomfield_FourAnal_2000}, multiplying the IC activations with $\exp(-\imagi 2\pi f T /2K)$.
This yields complex signals in the frequency band centered at $0$~Hz, the phase angles of which are plotted.

For multi-trial data, this results in two event-related potential (ERP) image plots \citep{JungMWTCS_ICA_1999,MakeigWJCTSC_JNeuSci_1999} showing event-related power and phase at each frequency $f$. 
For visual presentation, the trials are grayscale coded after sorting in order of ascending response time, followed by smoothing with a 30-trial wide rectangular window.
Response time in each trial is plotted superimposed on the data.
The time-courses of mean event-related power and intertrial coherence (ITC, \citet{makeig02:_dynam_brain_sourc_visual_evoked_respon}) may then be computed from the multi-trial data by averaging data from identical event-related time-windows across trials.

To visualize the complex component maps, the invariance of the source model \eqref{eq:8} with respect to rotation around the origin has to be taken into account.
Therefore, for each complex map $\veca_j(f)=[a_{1j}(f),\ldots,a_{Mj}(f)]^T$ any rotated version $c_j(f)\,\veca_j(f)$ thereof constitutes an equivalent map, with $c_j(f)$ an arbitrary unit-norm complex number.
For visualization we plot real-part, imaginary-part, magnitude and phase values of the equivalent map $\hat{\veca}_j(f) = c_j(f)\,\veca_j(f)$ for which the sum of the imaginary parts $\Im$ vanishes and the sum of the real parts $\Re$ is positive, i.e.,
\begin{gather}
  \label{eq:12}
  \sum_i \Im \left( \hat{a}_{ij}(f) \right) =  \Im \left( c_j(f) \sum_i a_{ij}(f) \right) = 0
  \qquad \text{and} \qquad
  \sum_i \Re \left( \hat{a}_{ij}(f) \right) > 0
  \\
  \Longrightarrow \qquad
  c_j(f) = \frac{\sum_i a_{ij}^{\ast}(f)}{\left| \sum_i a_{ij}(f) \right|} .
\end{gather}
A complex map $\hat{\veca}_j(f)$ whose elements $\hat{a}_{ij}(f)$ have negligible (near zero) imaginary part for all $i=1,\ldots,M$ indicates that the corresponding EEG process may represent activity of a highly synchronized generator ensemble, without phase shifts across the spatial extent of the source.
A non-negligible imaginary part is equivalent to phase-differences between distinct scalp electrode positions which may be elicited by spatio-temporal dynamics of the corresponding EEG process, e.g., spatial propagation of EEG activity.

\subsubsection{Degree of separation}
\label{sec:assess-qual-separ}

To quantify the degree of separation attained, we compute second and fourth order measures of statistical dependency.

Second order correlations are taken into account by computing, for each frequency $f$, the mean $\rho(f)$ of the absolute values of correlation-coefficients $\rho_{ij}(f)$ for all different component pairs $i \neq j$:
\begin{equation}
  \label{eq:13}
  \rho(f) = \frac{1}{N(N-1)} \sum_{i\neq j} \rho_{ij}(f),
\end{equation}
where the correlation-coefficients are defined as
\begin{align}
  \label{eq:5}
  \rho_{ij}(f) &= \left| \frac{\bra u_i(T,f) \, u_j^\ast(t,f) \ket_T - \mu_i(f)\,\mu_j^\ast(f)}{\sigma_i(f)\,\sigma_j(f)} \right|,\\
  \label{eq:16}
  \mu_i(f)     &= \bra u_i(T,f) \ket_T,\\
  \label{eq:19}
  \sigma_i(f)  &= \sqrt{ \bra \left|  u_i(T,f) - \mu_i(f)  \right|^2 \ket_T }.
\end{align}

$\rho_{ij}(f)$ vanishes for uncorrelated signals and acquires its maximum (one) only when signals $u_i(T,f)$ and $u_j(T,f)$ are proportional.
Since the measured signals $\vecx(T,f)$ are complex (except at 0~Hz and the Nyquist frequency), complete decorrelation may in general only be achieved by the fully-complex ICA algorithm (Eq.~\ref{eq:6}), whereas the real-map constrained-complex ICA algorithm (Eq.~\ref{eq:11}) and time-domain ICA will generally exhibit non-zero values of $\rho(f)$.

Second order decorrelation is not a sufficient condition for statistical independence. 
Therefore, we (partially) evaluate higher order statistical dependencies by computing the analog quantity $\rho^\prime(f)$ of the time-courses of squared amplitudes $|u_i(T,f)|^2$:
\begin{equation}
  \label{eq:14}
  \rho^\prime(f) = \frac{1}{N(N-1)} \sum_{i\neq j} \rho^\prime_{ij}(f),
\end{equation}
where
\begin{align}
  \label{eq:10}
  \rho^\prime_{ij}(f) &= \left| \frac{\bra |u_i(T,f)|^2 \, |u_j(T,f)|^2 \ket_T - \mu^\prime_i(f)\,\mu^\prime_j(f)}{\sigma^\prime_i(f)\,\sigma^\prime_j(f)} \right|,\\
  \label{eq:25}
  \mu^\prime_i(f)     &= \bra |u_i(T,f)|^2 \ket_T,\\
  \label{eq:26}
  \sigma^\prime_i(f)  &= \sqrt{ \bra \left(  |u_i(T,f)|^2 - \mu^\prime_i(f)  \right)^2 \ket_T }.
\end{align}

Eq.~\eqref{eq:10} measures statistical dependency of fourth order.
It can be interpreted as a modified and normalized variant of a fourth order cross-cumulant \citep{nikias93:_higher}.
Its value is zero for independent signals, non-zero for signals exhibiting correlated fluctuations in signal power, and maximum (one) only for signals with proportional squared-amplitude time-courses (regardless of phase).

\subsection{Corresponding components in distinct spectral bands}
\label{sec:align-source-comp}

The complex spectral-domain ICA algorithm described above produces separate sets of independent components for distinct and comparably narrow spectral bands.
Activity in some underlying EEG source domains might exhibit strictly narrow-band characteristics.
However, generator activity may also take place in a broader spectral range comprising contiguous or disconnected spectral bands.
Narrow-band ICA analysis does not take into account such links between bands, but separates the data into independent components ordered arbitrarily (e.g., by band-limited power) in each band.
Therefore, components that may account for activity within a single underlying EEG generator may be captured by components in multiple bands (with possibly distinct component numbers).
To obtain a full picture of the underlying EEG processes, it is desirable to identify and group together those components in different bands that likely represent activity of the same physiological source.

In this section, we present methods for identifying and clustering groups of similar components across frequencies.
The methods are based on appropriate measures of distance between pairs of component maps or component activations, respectively.
Matching component pairs are then identified using a standard optimal-assignment procedure.

\subsubsection{Distance between component maps}
\label{sec:comp-comp-maps}

Our definition of the distance between component maps is based on the euclidian distance $|\veca_i(f_1) - \veca_j(f_2)|$ of the complex vectors $\veca_i(f_1)$ and $\veca_j(f_2)$ representing two maps.
Since euclidian distance is not invariant with respect to arbitrary rescaling of the maps, it should be normalized.
The multiplication of one map with an arbitrary unit-norm complex number $c$, $|c|=1$, also alters the euclidian distance, although it results in an equivalent map.
Therefore, we define the map distance $d_\text{map}(i,f_1,j,f_2)$ of maps $\veca_i(f_1)$ and $\veca_j(f_2)$ as the rescaled minimal euclidian distance between the normalized maps,
\begin{equation}
  \label{eq:23}
  d_\text{map}(i,f_1,j,f_2) = 
      \frac{1}{\sqrt{2}}
      \underset{c}{\text{min}} 
      \left| c \, \frac{ \veca_i(f_1)}{|\veca_i(f_1)|} 
      - \frac{\veca_j(f_2)}{|\veca_j(f_2)|}
      \right|,
      \qquad |c|=1,
\end{equation}
which is written equivalently in terms of their innner-product as
\begin{gather}
  d_\text{map}(i,f_1,j,f_2) = 
      \underset{c}{\text{min}} 
      \sqrt{1-\left(
           \frac{\Re \left(c \, \veca_i^H(f_1) \, \veca_j(f_2)\right)}
                    {|\veca_i(f_1)|  \,|\veca_j(f_2)|}
        \right)^2}
      =
      \sqrt{1- \left( \frac{ | \veca_i^H(f_1) \, \veca_j(f_2) | }
                    {|\veca_i(f_1)|  \,|\veca_j(f_2)|} \right)^2
        }.
\end{gather}
The map distance measure attains its maximum (one) for orthogonal maps and its minimum (zero) only for equivalent maps.

\subsubsection{Distance between component activations}
\label{sec:comp-comp-activ}

We define the distance between complex component activations based on the correlation of signal-power time-courses at different frequencies.%
\footnote{Second order correlation drops off sharply with spectral difference because of the orthogonality of the Fourier basis and therefore is not appropriate for computing distances across different spectral bands.}
Between IC activations $u_i(T,f_1)$ and $u_j(T,f_2)$ at frequencies $f_1$ and $f_2$, respectively, the component activation distance $d_\text{act}(i,f_1,j,f_2)$ may be defined as
\begin{equation}
  \label{eq:24}
  d_\text{act}(i,f_1,j,f_2) = 1 - \rho^\prime_{ij}(f_1,f_2),
\end{equation}
where (analogous to Eq.~\ref{eq:10}), $\rho^\prime_{ij}(f_1,f_2)$ denotes the correlation-coefficient of the squared-amplitude time-courses $|u_i(T,f_1)|^2$ and $|u_j(T,f_2)|^2$,
\begin{equation}
  \label{eq:15}
  \rho^\prime_{ij}(f_1,f_2) = \left| \frac{\bra |u_i(T,f_1)|^2 \, |u_j(T,f_2)|^2 \ket_T - \mu^\prime_i(f_1)\,\mu^\prime_j(f_2)}{\sigma^\prime_i(f_1)\,\sigma^\prime_j(f_2)} \right|,
\end{equation}
with $\mu_i^\prime(f)$ and $\sigma_i^\prime(f)$ defined according to Eqs.~(~\ref{eq:25}) and (\ref{eq:26}), respectively.
By this measure, independent signals have maximal distance (one), whereas signals with highly correlated fluctuations in signal power have distance near minimum (zero).
Related changes in signal power in different frequency bands may be exhibited by EEG generators with activity in both bands, since modulation of generator activity---induced, e.g., by experimental events or common modulatory processes---may result in synchronous amplitude changes (in the same or different direction) in the participating bands.

\subsubsection{Assigning best-matching component pairs}
\label{sec:hungarian-method}

Based on the distance measures described in sections \ref{sec:comp-comp-maps} and \ref{sec:comp-comp-activ}, we define the set of pairs of best-matching components to be that which minimizes the average distance between the pairs.

Consider a given pair of frequencies $(f_1, f_2)$ and a chosen distance measure $d(i,f_1,j,f_2)$ (either map distance $d_\text{map}$ or activation distance $d_\text{act}$).
Assigning best-matching component pairs is equivalent to finding the permutation $\pi(i)$, $i=1,\ldots,N$, that assigns component $i$ at frequency $f_1$ to component $j=\pi(i)$ at frequency $f_2$ such that the mean distance across all pairs, $\sum_i d(i,f_1,\pi(i),f_2)/N$, is minimized:
\begin{align}
  \label{eq:22}
  \pi (\cdot) &= \underset{\pi(\cdot)}{\text{argmin}} \sum_i d(i,f_1,\pi(i),f_2),\\
  \label{eq:17}
  D(f_1,f_2)  &= \underset{\pi(\cdot)}{\text{min}} \frac{1}{N} \sum_i d(i,f_1,\pi(i),f_2).
\end{align}
Determining $\pi(i)$ given the matrix of distances $d(i,f_1,j,f_2)$ between all pairs $(i,j)$ is known as the `assignment problem'.
A classic algorithm for solving this problem is the Hungarian method \citep{Kuhn_NavResLogist_1955} which we use here following the suggestion of \citet{Enghoff_Thesis_1999}.

The minimal mean distance $D(f_1,f_2)$ is a global measure of the distance between the sets of components at frequencies $f_1$ and $f_2$.
For equal frequencies, $f_1=f_2$, $D(f_1,f_2)$ always attains its minimum (zero), and the permutation becomes the identity, $\pi(i)=i$.
If the components at frequency $f_1$ are identical to the components at frequency $f_2$, but occur in a different order, then $D(f_1,f_2)$ is also zero and $\pi(i)$ corresponds to the permutation of order.
If some components are identical at both frequencies, whereas the remaining components exhibit maximum distance to all other components, then $D(f_1,f_2)$ corresponds to the fraction of non-identical components.
For the realistic case of few components being reproduced exactly across frequencies and many components matching similar but not identical components at other frequencies, $D(f_1,f_2)$ attains values between zero and one, indicating the degree of average similarity of the best-matching component pairs.

\subsection{Time-domain ICA}
\label{sec:time-domain-ica}

We analyze separation results from time-domain ICA using similar methods as those presented in sections 2.3.3 and 2.4 for the analysis of frequency-domain ICA.
Time-domain infomax ICA is applied to the time-domain signals $x_i(t)$, resulting in a single separating matrix $\matW$.
The corresponding components maps are given by the columns $\veca_j$ of the mixing matrix $\matA=\matW^{-1}$.
We obtain frequency-specific unmixed signals by applying $\matW$ to the spectral transforms of the sources, yielding complex separated signals $\vecu(T,f) = \matW \, \vecx(T,f)$, from which we compute the measures for the quality of separation (cf.\ section \ref{sec:assess-qual-separ}).
Distances between time-domain and spectral-domain components are obtained based on the methods presented in section \ref{sec:align-source-comp}.
The distance $d_{\text{map}}(i,j,f)$ between time-domain ICA maps $\veca_i$ and spectral-domain ICA maps $\veca_j^\prime(f)$ is computed analogously to Eq.~(\ref{eq:23}).
Similarly, IC activations  obtained with time-domain and frequency-domain ICA are compared by computing the distance $d_{\text{act}}(i,j,f)$ in analogy to Eq.~(\ref{eq:24}).
We then assign best-matching component pairs using the method presented in section \ref{sec:hungarian-method}.

\section{Results}
\label{sec:results}

In this section, we present results from the analysis of a visual spatial selective attention experiment where the subject attended one out of five indicated locations on a screen while fixating a central cross, and was asked to respond by a button press as quickly as possible each time a target stimulus appeared in the attended location.
For details of the experiment, see \citet{MakeigWTJCS_PhilTransRoySocLonB_1999}.
Included in the analysis were 582 trials from target stimulus epochs collected from one subject.
Each epoch was $1$ s long, beginning at $100$ ms before stimulus onset at $t=0$ ms.

The data were recorded from 31 EEG electrodes (each refered to the right mastoid) at a sampling rate of 256~Hz and decomposed into $101$ equidistantly spaced spectral bands with center frequencies from $0.0$~Hz (DC) to $50.0$~Hz in $0.5$-Hz steps.
Decomposition was performed by short-time discrete Fourier transformation with a hanning window of length $50$ samples, corresponding to a spectral resolution of $5.12$~Hz (defined as half width at half maximum), and a window shift of $1$ sample between successive analysis windows.
This yielded 207 short-time spectra for each trial derived from analysis windows centered at times between $1.6$ ms and $806.3$ ms following stimulus presentations in $3.9$ ms steps.

To decompose the data into independent components, the 582 trials were concatenated, resulting for each spectral band, $f=0,\ldots,101$, and channel, $i=1,\ldots,31$, in frames $T=1,\ldots,207 \times 582 = 120474$.
No pre-training sphering of the data was performed.
The separating matrix $\matW(f)$ was initialized with the identity matrix for all spectral bands.
We used the logistic non-linearity (Eq.~\ref{eq:9}), computed the gradients (Eq.~\ref{eq:6}) and (Eq.~\ref{eq:11}), respectively, at each iteration step from 10 randomly chosen data points, and lowered the learning rate of the gradient ascent procedure successively.
Optimization of $\matW(f)$ was halted when the total weight-change induced by one sweep through the whole data was smaller than $10^{-6}$ relative to the Frobenius norm of the weight-matrix.

The dataset was decomposed using both the fully-complex (Eq.\ \ref{eq:6}) and the real-map constrained-complex (Eq.\ \ref{eq:11}) algorithms.
For comparison, the same dataset was also decomposed using time-domain infomax ICA applied to the time-domain data $x_i(t)$; the obtained single real separating matrix was then applied to the spectral-domain data $\vecx(T,f)$ as described in section \ref{sec:time-domain-ica}.

\subsection{Kurtosis}
\label{sec:kurtosis}

\begin{figure}[t]
  \centering
  \includegraphics[width=0.45\linewidth,keepaspectratio]{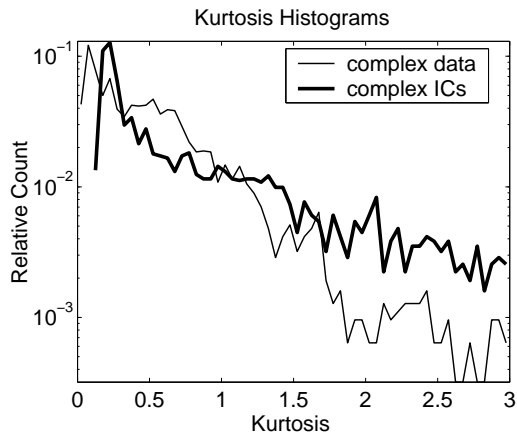}
  \caption{Histograms for estimated kurtosis of complex spectral-domain electrode signals (thin line) and independent component activations (thick line). Each histogram based on 3131 kurtosis estimates (see text), $44$ bins of width $0.05$ in the interval from $0$ to $3$.}
  \label{fig:kurtosis}
\end{figure}

To test the assumption of super-Gaussian source distributions, we used kurtosis to assess deviations from a Gaussian distribution.
Kurtosis estimates were computed for spectral-domain data as
\begin{equation}
  \label{eq:21}
  \text{kurt}(z) =  \bra |z|^4 \ket - 2 \left( \bra |z|^2 \ket \right)^2 - | \bra zz \ket |^2
\end{equation}
assuming a zero-mean, unit-variance complex random variable $z$ \citep{HyvarinenKO_2001}.
The Kurtosis $\text{kurt}(z)$ vanishes for a Gaussian distribution and attains positive and negative values for super- and sub-Gaussian distributions, respectively.

Kurtosis of the spectral-domain electrode signals $x_i(T,f)$ was computed individually for each channel $i$ at every frequency $f$, yielding $31 \times 101=3131$ kurtosis estimates, each based on all $120474$ complex data frames.
All of the $3131$ channel-frequency kurtosis estimates showed a super-Gaussian distribution with minimum $0.02$, maximum $23.45$ and median $0.43$.
A histogram of the kurtosis values is displayed in Fig.~\ref{fig:kurtosis} (thin line).

Analogously, we computed the same number of kurtosis estimates for the IC activations $u_j(T,f)$ obtained with the fully-complex ICA algorithm.
The median kurtosis increased to $0.55$ and only super-Gaussian distributions in the range $[0.10, 386.79]$ were found.
Their histogram is shown in Fig.~\ref{fig:kurtosis} (thick line).

These results support our choice of source model, and indicate that only a small advantage might be expected by allowing the source distributions to include sub-Gaussian sources.
Therefore, we did not consider the possibility of sub-Gaussian sources further.

\subsection{Degree of Separation}
\label{sec:quality-separation}

\begin{figure}[t]
  \centering
  \includegraphics[width=0.45\linewidth,keepaspectratio]{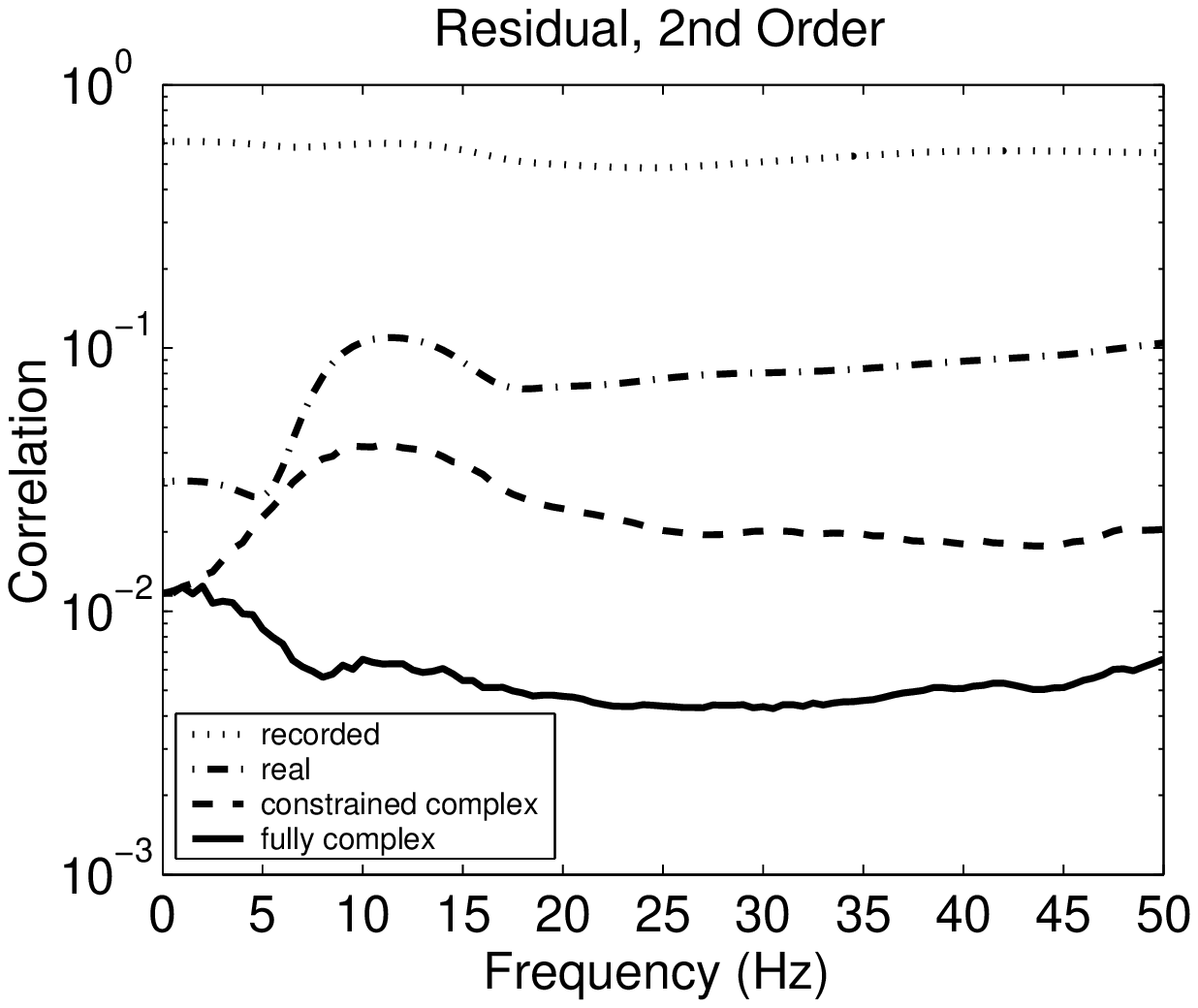}
  \includegraphics[width=0.45\linewidth,keepaspectratio]{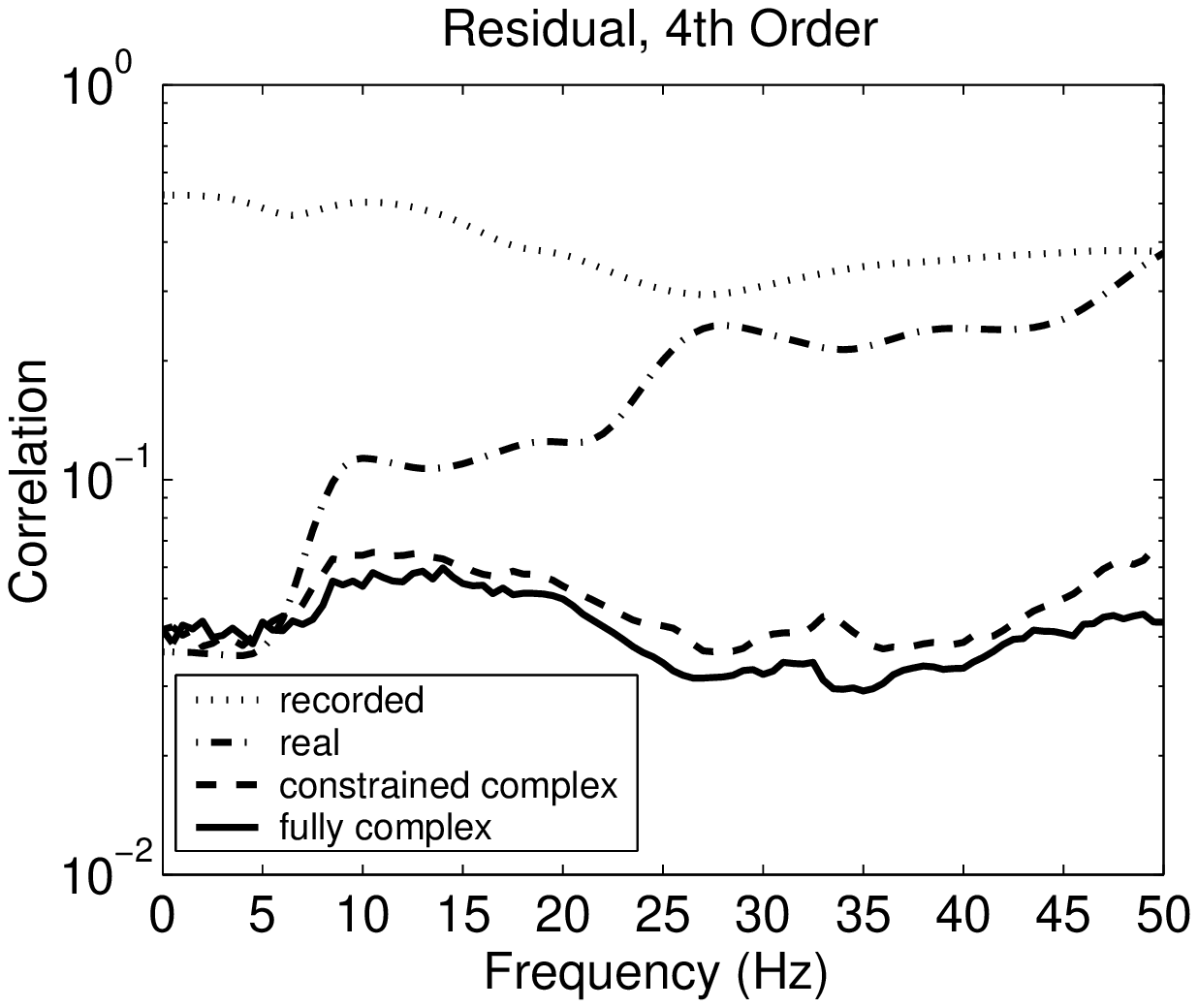}
  \caption{Residual statistical dependencies evaluated using second order (left panel) and fourth order (right panel) measures at frequency bands between 0~Hz and 50~Hz. Residuals for the recorded electrode signals (dotted), signal separation obtained from real time-domain infomax ICA (dash-dotted), real-map constrained-complex spectral-domain ICA (dashed), and fully-complex spectral-domain ICA (solid).}
  \label{fig:sepquality}
\end{figure}

To assess the degree of separation achieved by the different ICA algorithms, we computed residual statistical dependencies using the second order (Eq.~\ref{eq:13}) and fourth order (Eq.~\ref{eq:14}) statistics described in section~\ref{sec:assess-qual-separ}.
Results are displayed in Fig.~\ref{fig:sepquality} for the recorded electrode signals and for the separations into sources obtained from real time-domain infomax ICA, real-map constrained-complex and fully-complex spectral-domain ICA.
For both measures and all frequencies, fully-complex ICA achieved the lowest levels of residual dependencies.
Real-map constrained results exhibited comparably higher residuals, and time-domain infomax ICA still higher levels.

The residual second order correlations exhibited by fully-complex ICA were---with the exception of very low frequencies---about one order of magnitude lower than those attained by time-domain ICA, and below half of those achieved by real-map constrained-complex ICA.
This result may largely be explained by the higher number of degrees of freedom of the complex ICA algorithms that model the superposition within each frequency band with a different mixing matrix, whereas time-domain ICA uses a single matrix for all frequencies.
Fully-complex ICA achieved the lowest levels of residual correlation since it is the only algorithm that models superposition using a different complex matrix for every frequency, which in general is necessary to decorrelate complex input signals.
In the 0-Hz frequency band, the frequency-domain electrode signals are real, which explains the similar performance of the real-map constrained- and fully-complex algorithms at the lower end of the spectral range.

The residual fourth order correlations showed a smaller difference between the real-map constrained and fully-complex ICA algorithms, the latter exhibiting slightly lower residual dependencies for all but very low frequencies.
Remarkably, there was almost no difference in fourth order correlations between the three algorithms in the range from 0~Hz to approximately 6~Hz, which may be due to high power of the signals in this range.
Therefore, time-domain ICA may be best capable of separating signals in this spectral region.
Between about 6~Hz and 50~Hz, the residual fourth order correlations of time-domain ICA showed large fluctuations---near 27~Hz and 50~Hz component independence was close to that of the recorded signals.

These findings indicate that additional degrees of freedom of the spectral-domain convolutive mixing model (compared to the instantaneous mixing model) enable it to produce components with a higher degree of signal separation.
If the underlying EEG processes had wide-band characteristics and no spatio-temporal dynamics, it would have been expected that all three algorithms performed equally well.
Therefore, the assumption of a fully-complex frequency-specific mixing model appears to be supported by the resulting lower residual dependencies.

\subsection{Distance between component maps}
\label{sec:clust-comp-maps}

\begin{figure}[t]
  \centering
  \includegraphics[width=0.45\linewidth,keepaspectratio]{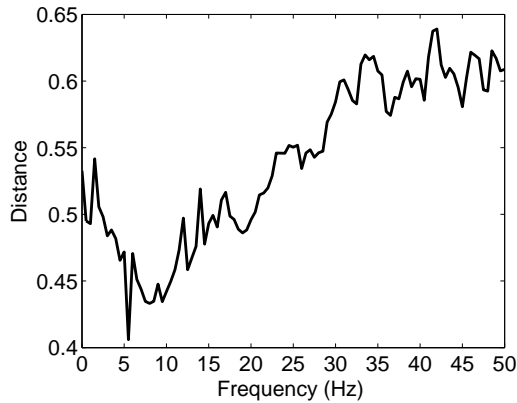}
  \caption{
    Mean distance between the component maps obtained by time-domain infomax ICA and best-matching frequency-specific component maps of real-map constrained-complex ICA.
    Abscissa: frequency of spectral-domain component. Ordinate: mean distance to time-domain ICA map.}
  \label{fig:inst_map_dist}
\end{figure}

We further compared time-domain ICA and real-map constrained-complex ICA by computing, for every frequency $f=1,\ldots,101$, the distance $d_\text{map}(i,j,f)$ between the $i$-th component map of time-domain ICA and the $j$-th component map of complex ICA at frequency $f$, see section \ref{sec:time-domain-ica}.
Best-matching component maps were assigned for each $f$ using the assignment method described in section \ref{sec:hungarian-method}, yielding a minimal mean distance $D(f)$ (analogous to Eq.~\ref{eq:17}), which is shown in Fig.~\ref{fig:inst_map_dist}.

Across all frequencies, the distance between component maps obtained by time-domain ICA and by constrained-complex spectral-domain ICA is at least $0.4$.
Largest distances are exhibited at frequencies of 30~Hz or higher, while the maps show closest resemblance around a minimum in the 5-Hz to 10-Hz range.
In conjunction with the results from section \ref{sec:quality-separation}, this may serve as a further indication that separation of EEG data by time-domain ICA may be dominated by low-frequency activity.

\subsection{Distance between component activations}
\label{sec:dist-comp-activ}

\begin{figure*}[t]
  \centering
  \includegraphics[width=0.95\linewidth]{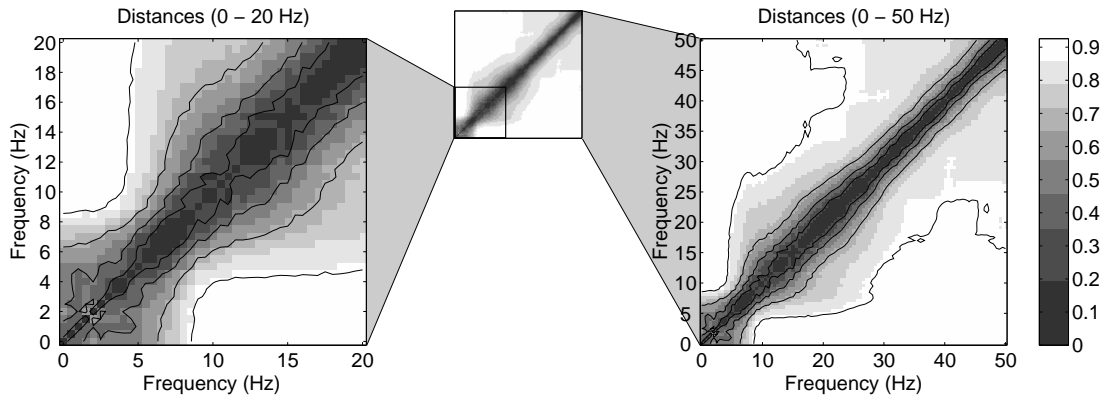}
  \caption{Minimal mean distances $D_\text{act}(f_1,f_2)$ computed from component activation functions obtained with the fully-complex ICA algorithm in 101 frequency bands of width 5.12~Hz, spaced equidistantly between 0~Hz and 50~Hz in 0.5-Hz increments. Right: Distances for all best-matching component pairs of different frequencies. Left: Enlarged view of the 0-Hz to 20-Hz range.}
  \label{fig:comp_dist}
\end{figure*}

Distances between component activation time-courses $u_i(T,f_1)$ and $u_j(T,f_2)$ were computed for the fully-complex ICA separation according to Eq.~(\ref{eq:24}) for all possible combinations of $(i,f_1,j,f_2)$.
Best-matching components were assigned for each pair of frequencies $(f_1,f_2)$ using the method presented in section \ref{sec:hungarian-method}, yielding one minimal mean distance $D_\text{act}(f_1,f_2)$ for every frequency pair.
The distances between all frequency pairs are visualized in Fig.~\ref{fig:comp_dist} (right panel).
Note that the level of detail available in the visualized spectral features is in principle limited by the spectral resolution ($5.12$~Hz) of the windowing function employed in the time-frequency transformation.

The distance matrix is dominated by values on the diagonal as expected from the bandwidth of the spectral decomposition.
For larger spectral distances (away from the diagonal) several nearly rectangular distance patterns emerge, that deviate from the diagonal structure as expected in the absence of spectral clusters.
Qualitatively, we may identify three spectral blocks, extending roughly from 0~Hz to 8~Hz (corresponding to delta and theta bands), from 8~Hz to 30~Hz (alpha and beta bands), and from 30~Hz to at least 50~Hz (gamma band).

Although the exact borders and shapes of the spectral clusters cannot be identified in the figure, the existence of spectral structure may reflect physiological processes that extend over some spectral range and thereby induce independent components with similarities across frequencies.
Thus, the complex spectral-domain ICA method may serve as the starting point for extracting components with physiological relevance from EEG data.
Whereas our present analysis of the clusters is based on the qualitative interpretation of the average component distances across frequencies, further analysis may employ quantitative clustering methods on individual (unaveraged) component distances and should thereby produce a more detailed picture of component similarities across frequencies.

\subsection{Examples of maps and activations}
\label{sec:exampl-maps-activ}

\begin{figure*}[p]
  \centering
  \includegraphics[width=0.25\linewidth,keepaspectratio]{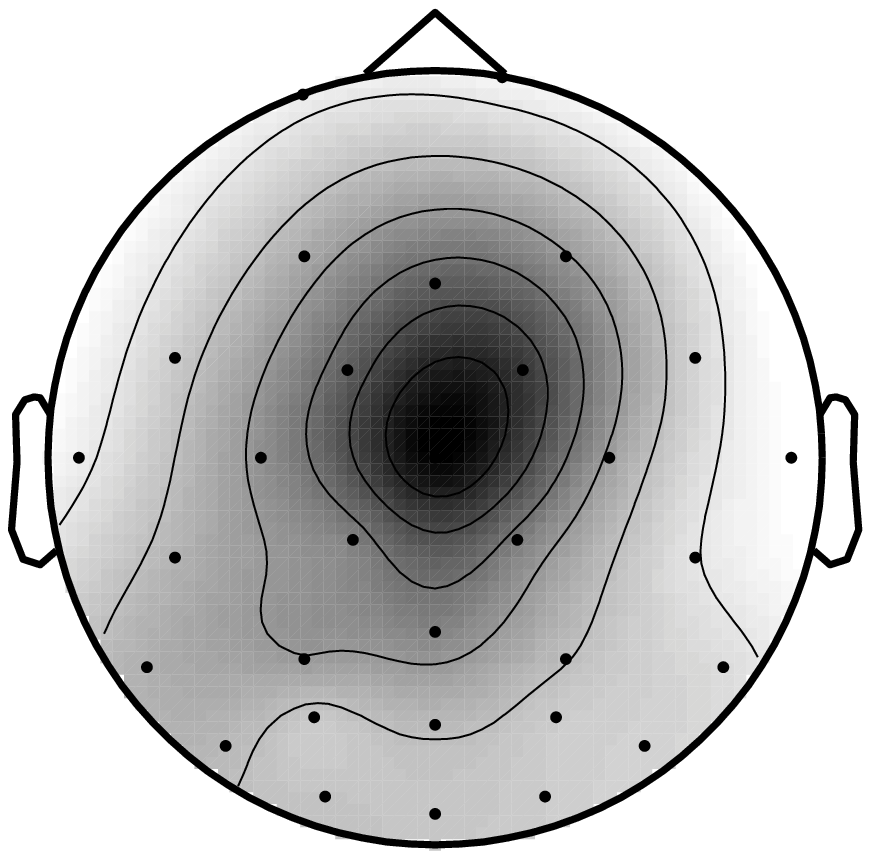}
  \hspace{0.04\linewidth}
  \includegraphics[width=0.33\linewidth,keepaspectratio]{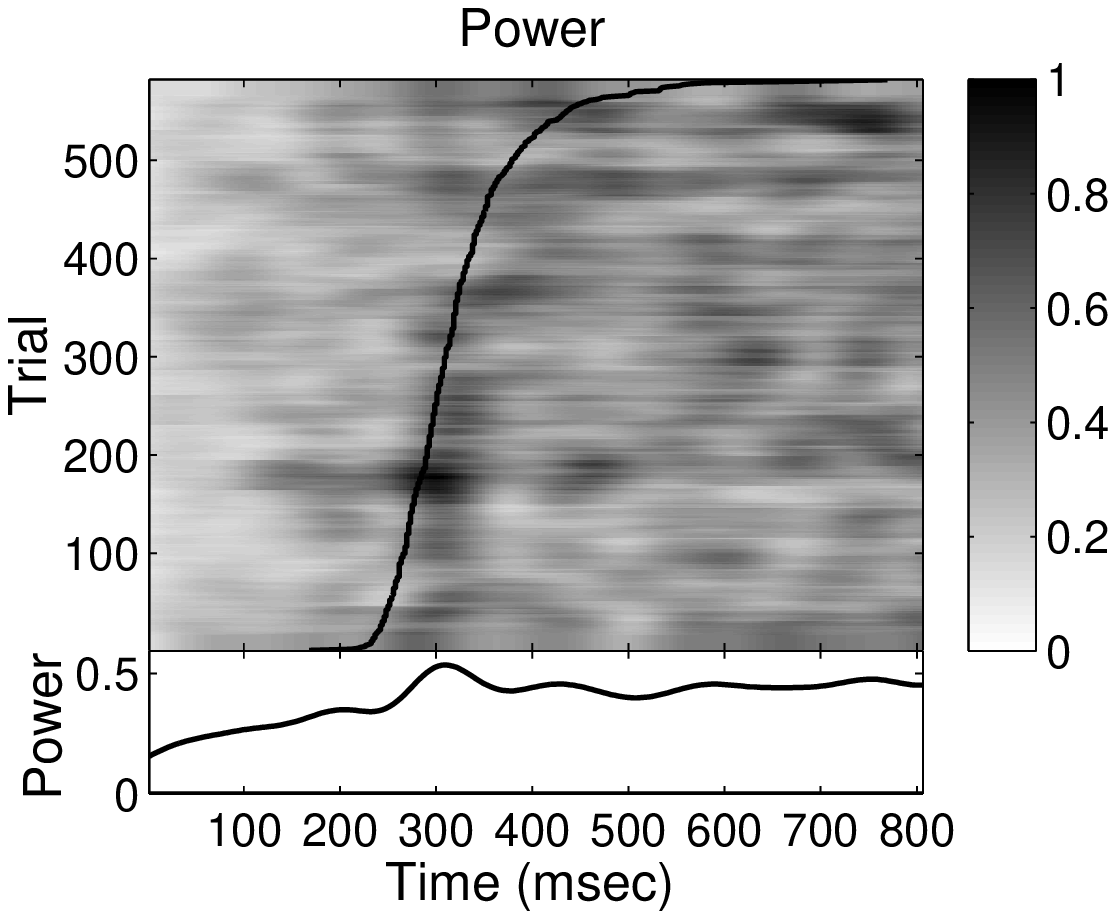}
  \includegraphics[width=0.33\linewidth,keepaspectratio]{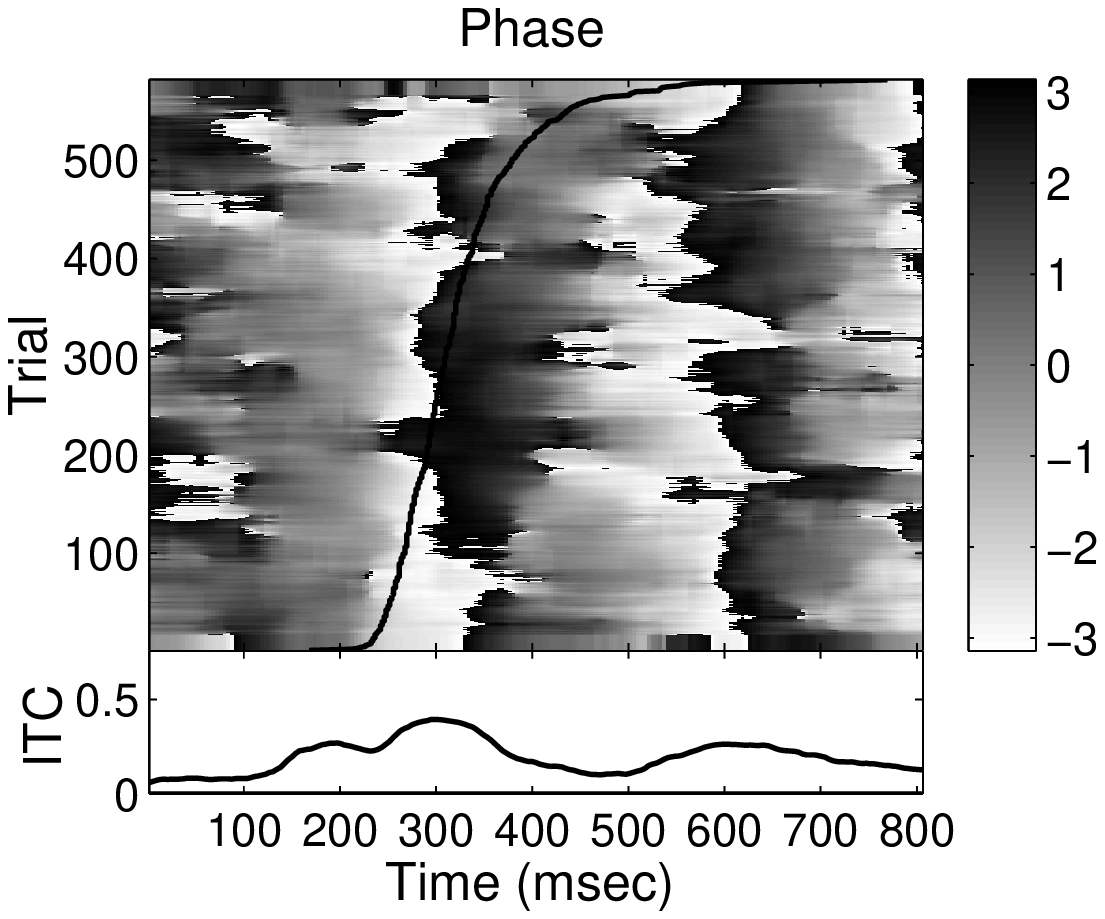}
  \caption{Independent component at 5~Hz obtained from standard time-domain infomax ICA. Left: Scalp map. Middle: ERP-image of 5-Hz power. Right: ERP-image of complex-demodulated 5-Hz phase. Response times superimposed on data. Lower panels: Mean time-courses of event-related 5-Hz power (middle) and 5-Hz intertrial coherence (ITC, right).}
  \label{fig:infomaxICA}
\end{figure*}

\begin{figure*}[p]
  \centering
  \includegraphics[width=0.25\linewidth,keepaspectratio]{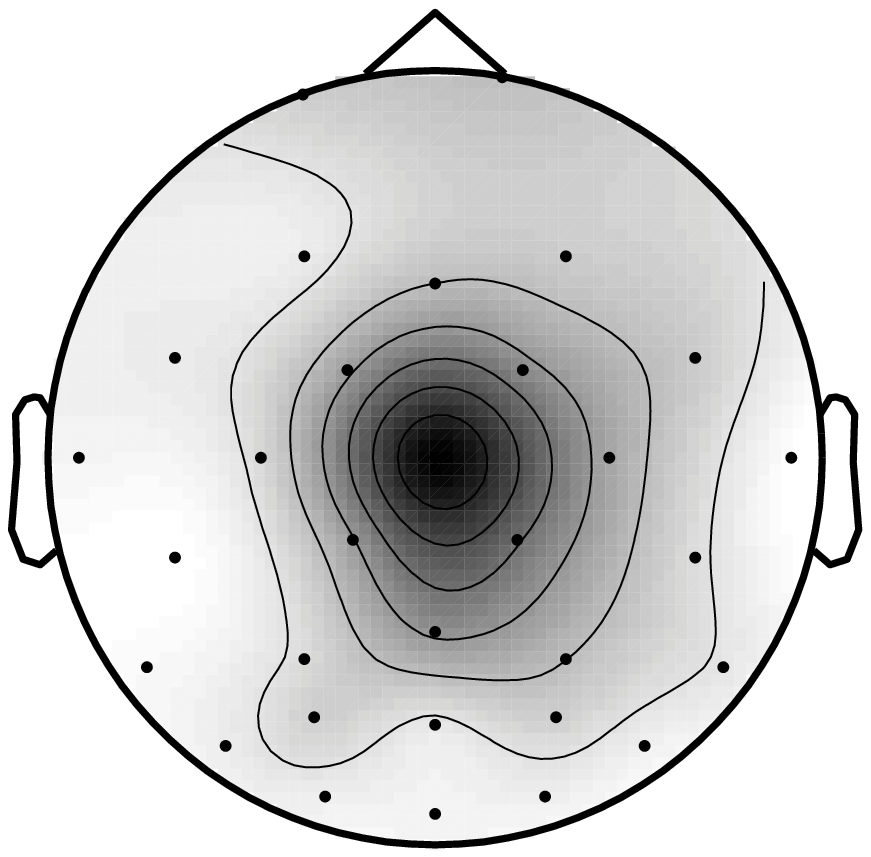}
  \hspace{0.04\linewidth}
  \includegraphics[width=0.33\linewidth,keepaspectratio]{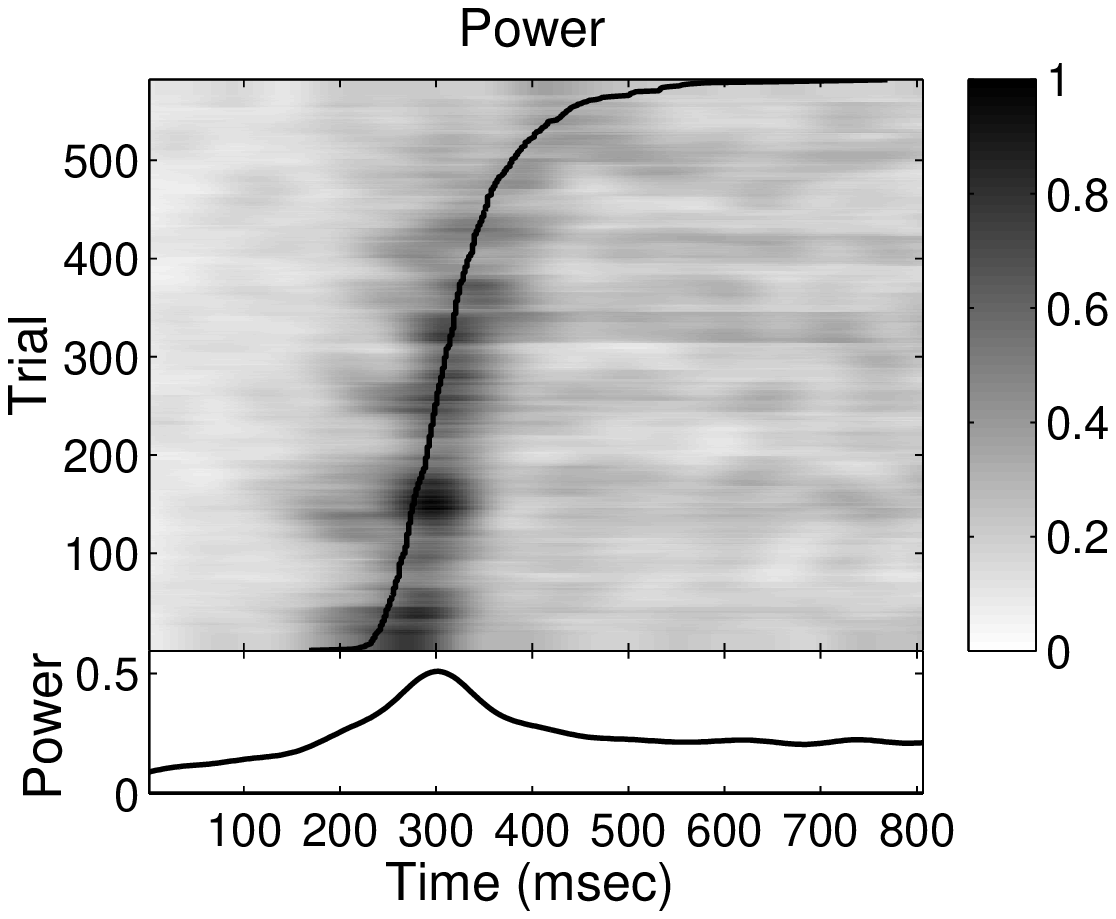}
  \includegraphics[width=0.33\linewidth,keepaspectratio]{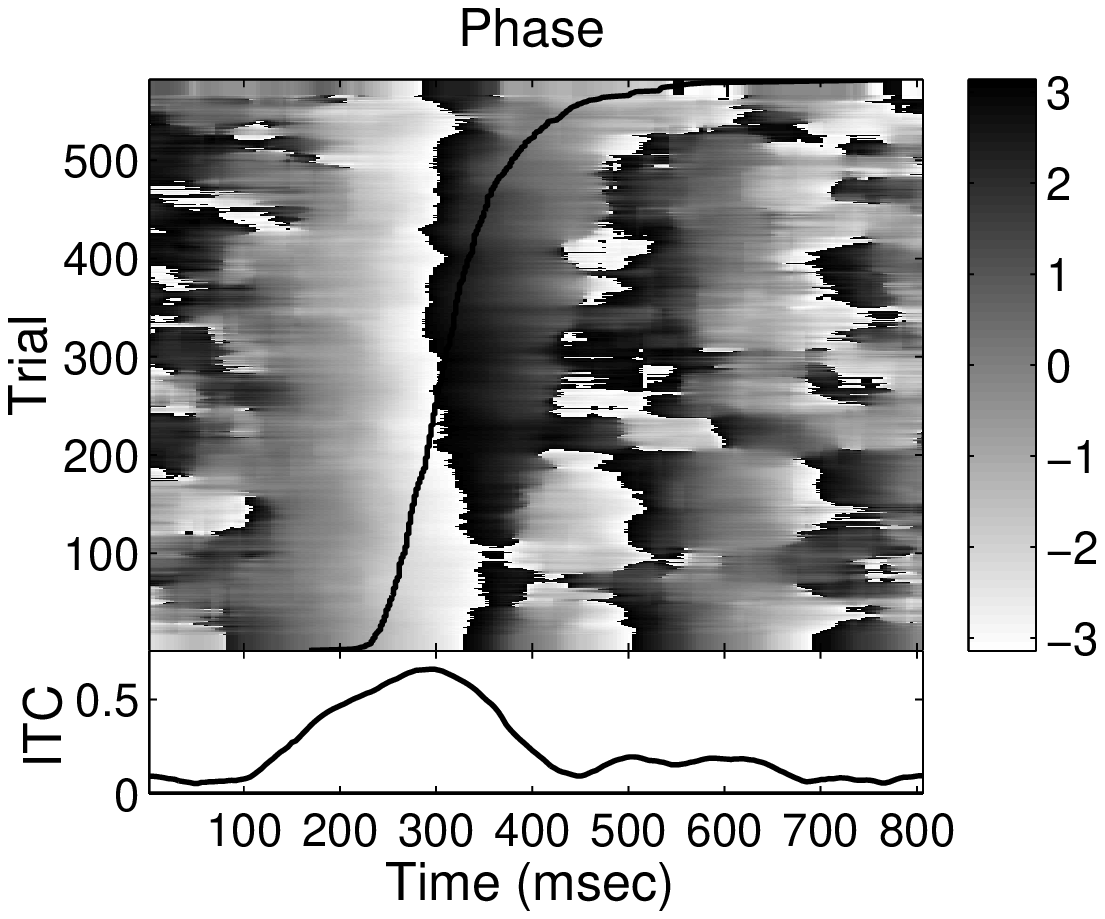}
  \caption{Independent component at 5~Hz obtained from real-map constrained-complex spectral-domain ICA. Same dataset as Fig.\ \ref{fig:infomaxICA}. Left: Scalp map. Middle: ERP-image of 5-Hz power. Right: ERP-image of complex-demodulated 5-Hz phase. Response time and lower panels analogous to Fig.\ \ref{fig:infomaxICA}.}
  \label{fig:realspecICA}
\end{figure*}

\begin{figure*}[p]
  \centering
  \includegraphics[width=0.19\linewidth,keepaspectratio]{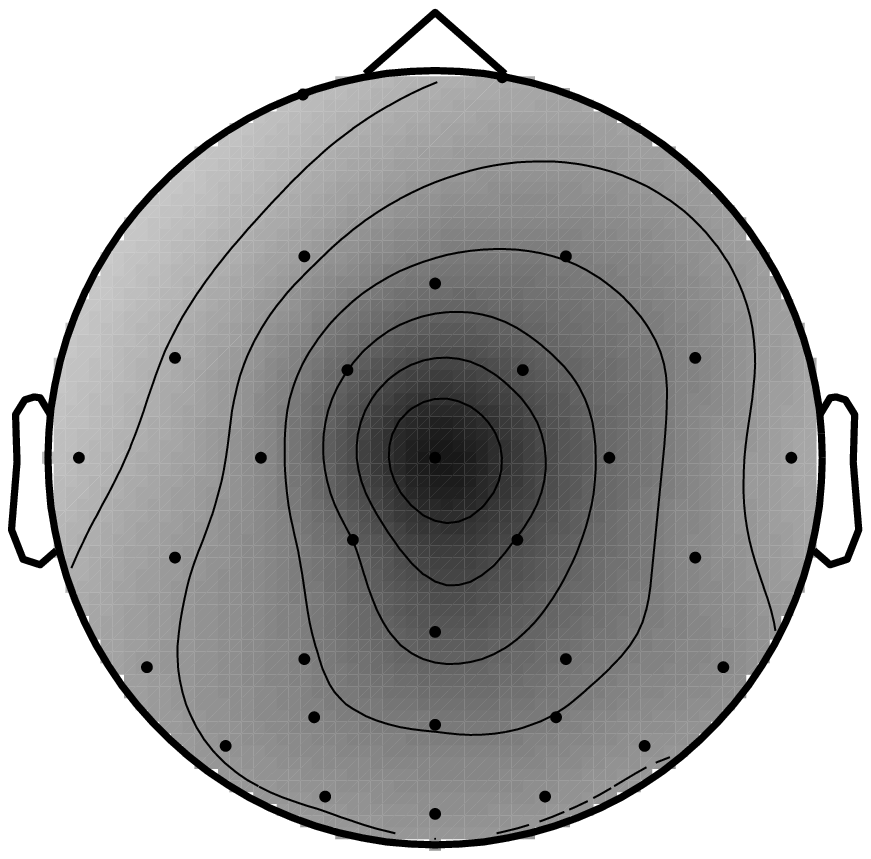}
  \includegraphics[width=0.19\linewidth,keepaspectratio]{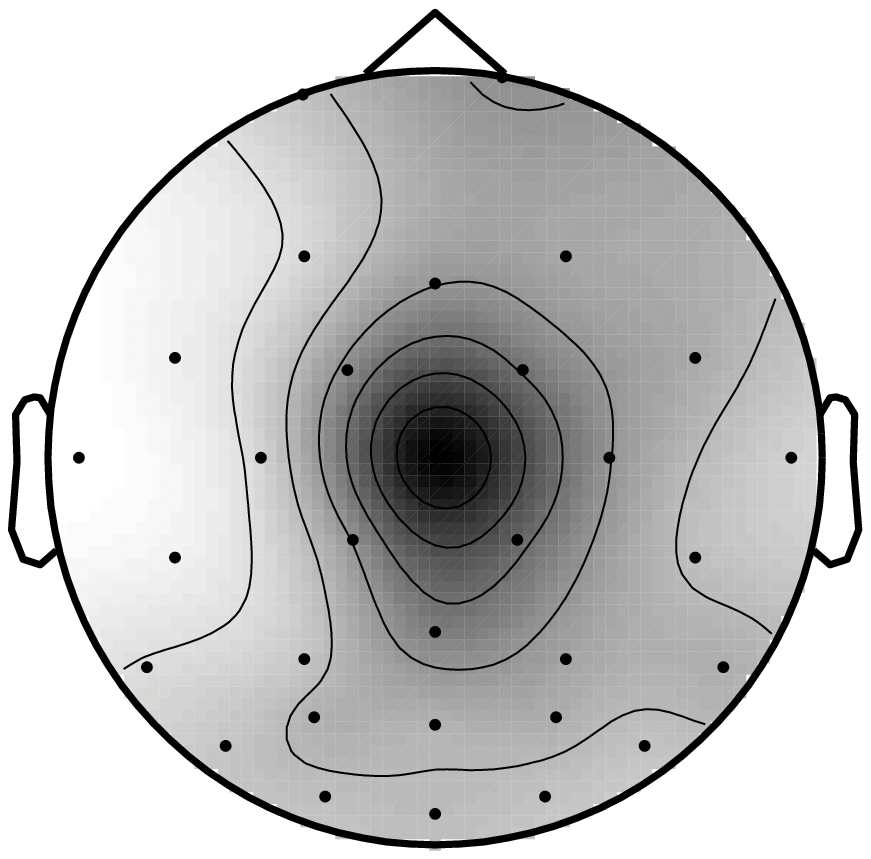}
  \hspace{0.03\linewidth}
  \includegraphics[width=0.27\linewidth,keepaspectratio]{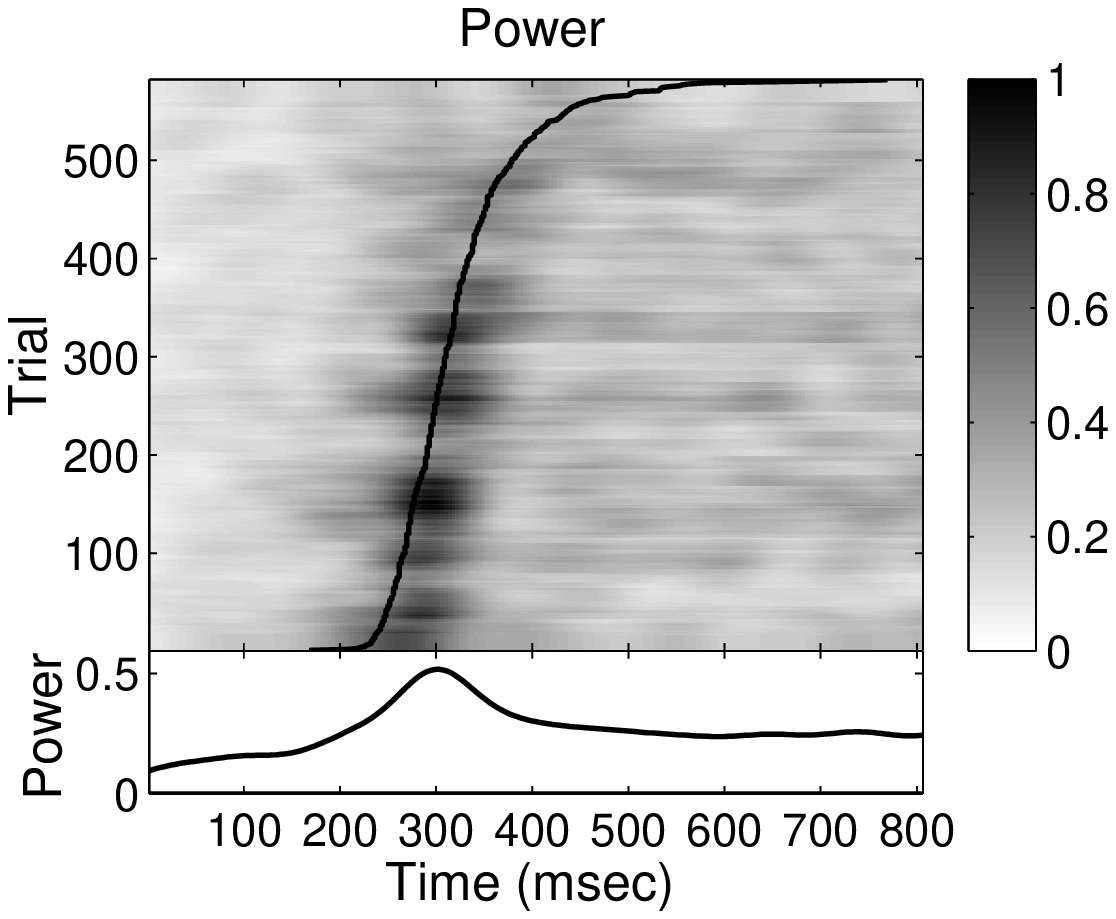}
  \includegraphics[width=0.27\linewidth,keepaspectratio]{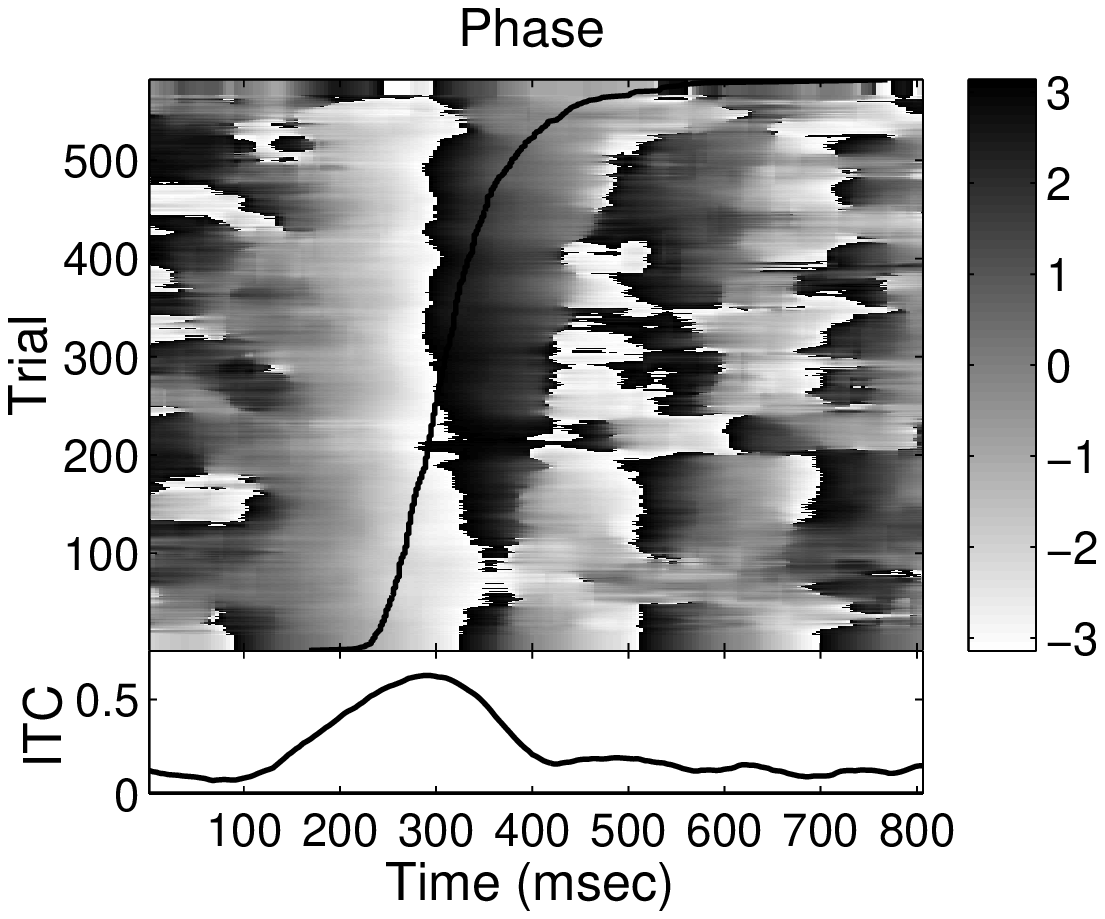}
  \caption{Independent component at 5~Hz obtained from fully-complex spectral-domain ICA. Same dataset as Figs.\ \ref{fig:infomaxICA} and \ref{fig:realspecICA}. From left to right: Real and imaginary part of the complex scalp map, respectively; ERP-images of 5-Hz power and complex-demodulated 5-Hz phase of the complex IC activation time-courses, respectively. Response time and lower panels analogous to Fig.\ \ref{fig:infomaxICA}.}
  \label{fig:cplxspecICA}
\end{figure*}

\begin{figure*}[t]
  \centering
  \begin{tabular}[h]{ccccc}
    10 Hz & 15 Hz & 20 Hz & 25 Hz & 30 Hz \\
    \includegraphics[width=0.177\linewidth,keepaspectratio]{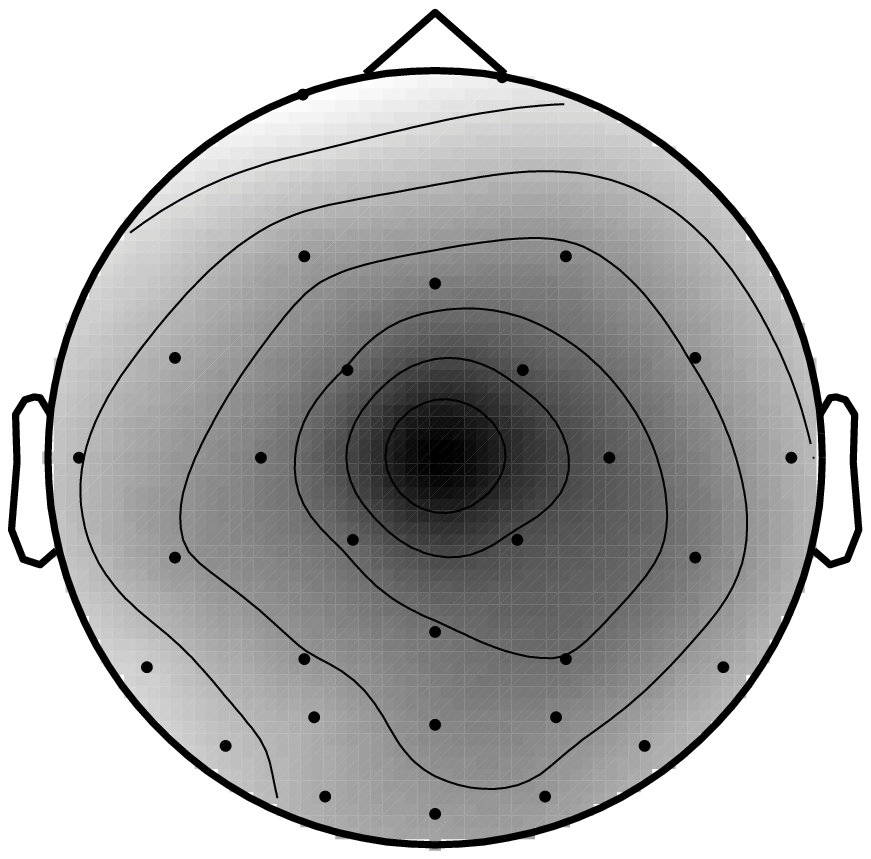}&
    \includegraphics[width=0.177\linewidth,keepaspectratio]{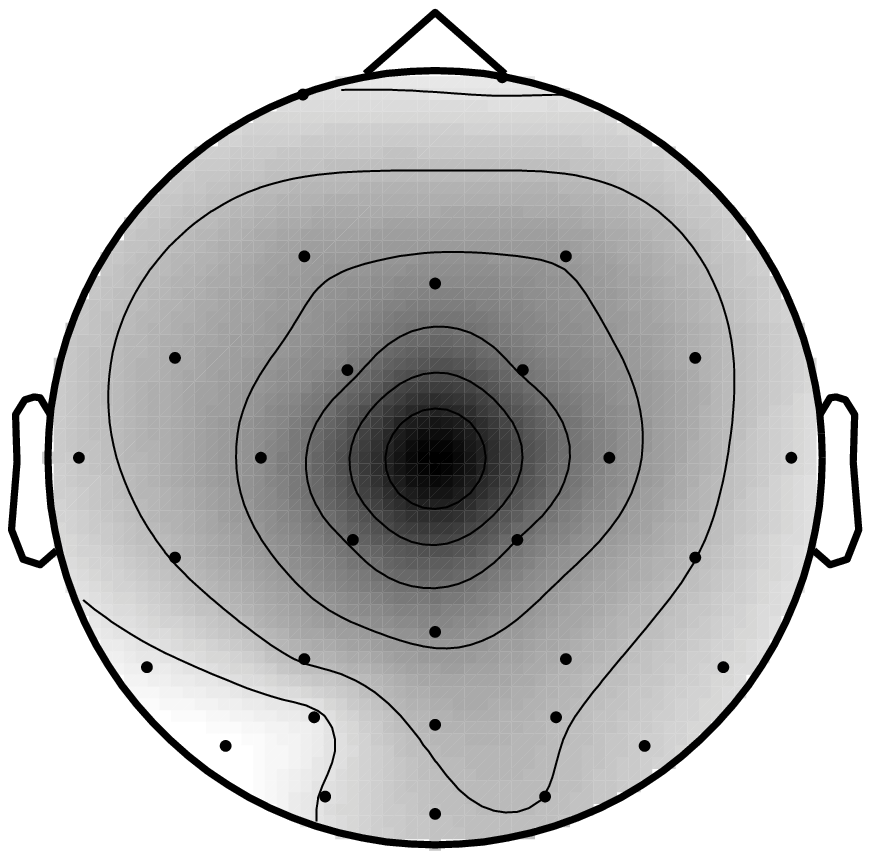}&
    \includegraphics[width=0.177\linewidth,keepaspectratio]{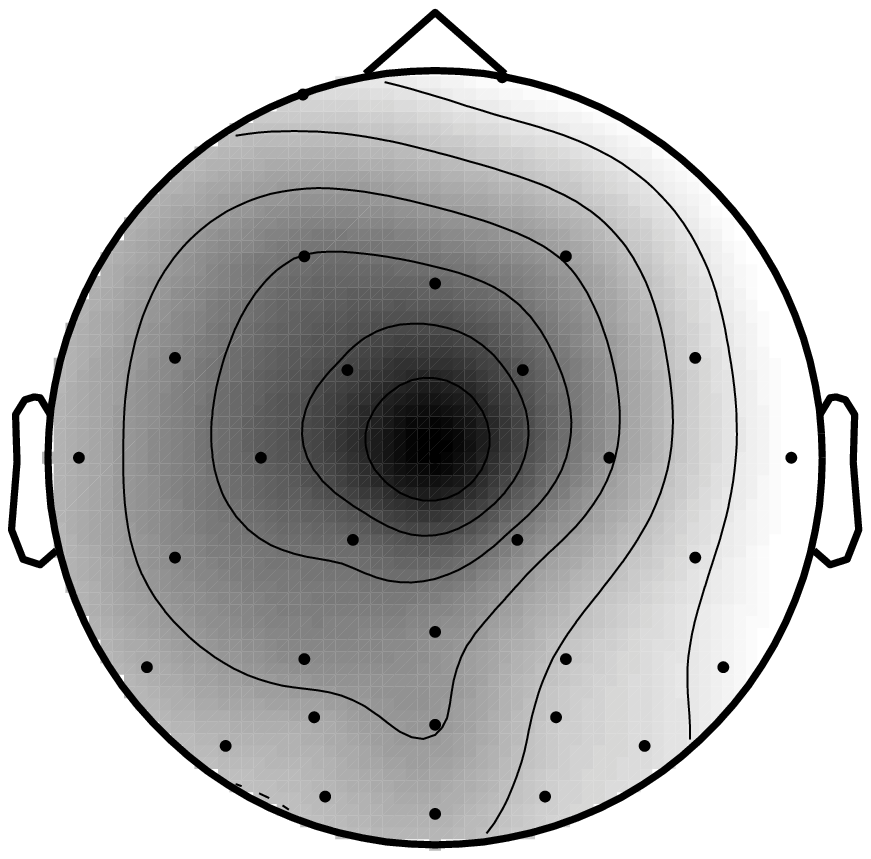}&
    \includegraphics[width=0.177\linewidth,keepaspectratio]{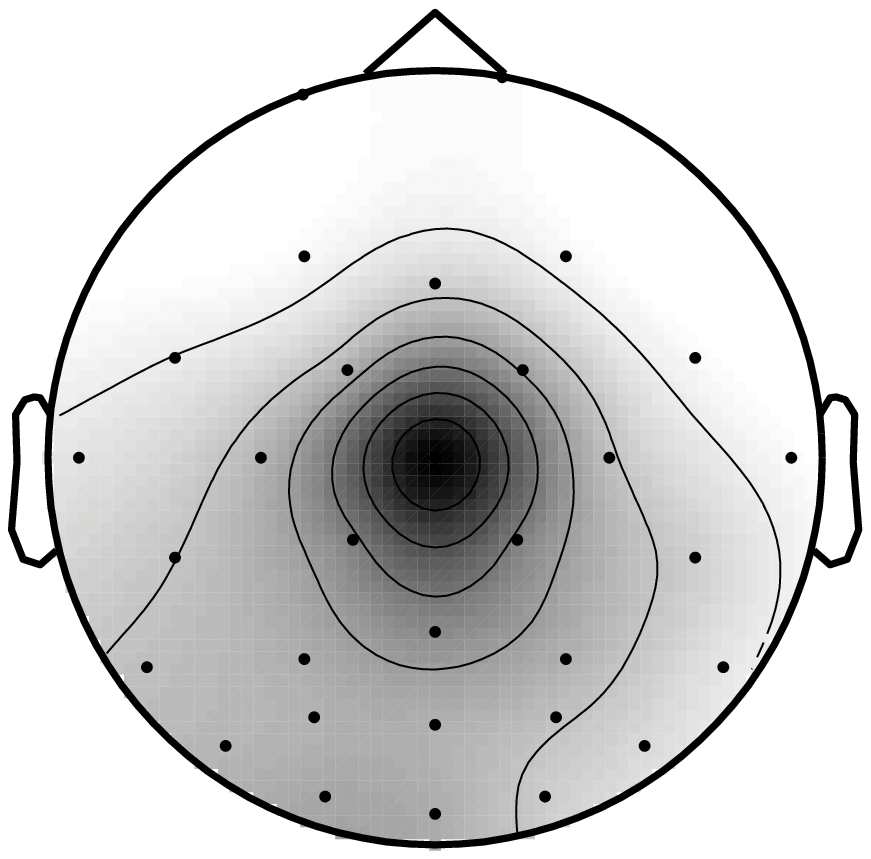}&
    \includegraphics[width=0.177\linewidth,keepaspectratio]{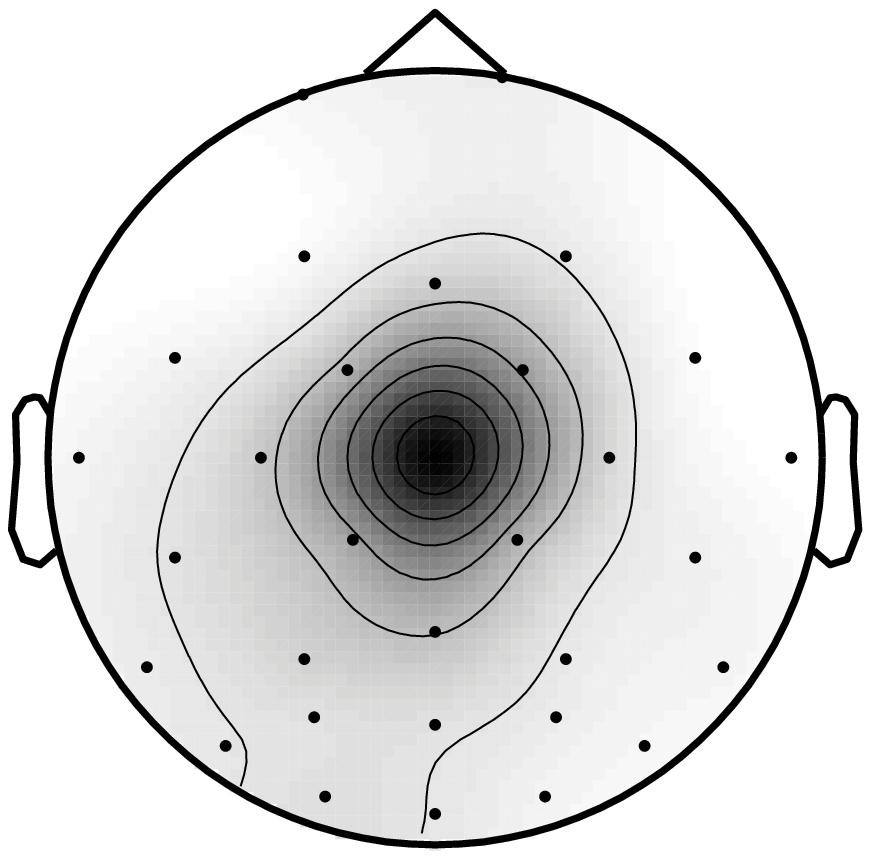}
  \end{tabular}
  \caption{Magnitude maps of complex independent components obtained using the fully-complex spectral-domain ICA algorithm at five frequency bands, same dataset as Figs.\ \ref{fig:infomaxICA}-\ref{fig:cplxspecICA}.}
  \label{fig:map_variability}
\end{figure*}

A large number of independent component maps and activations were obtained for different frequency bands.
Relevance of the components may be assessed based on the compatibility of component maps and activations with known EEG physiology, experimental design and subject behavior.
We show here one set of components whose central-midline projections are similar to EEG activity associated with orienting to novel stimuli \citep{CourchesneHG_ElEncClinNeuPhys_1975}.
The response of these components to stimulus presentation is most marked in the 5-Hz band and shows a clear relation to subject behavior.

Figs.\ \ref{fig:infomaxICA}--\ref{fig:cplxspecICA} illustrate differences between the real infomax, real-map constrained-complex infomax and fully-complex infomax ICs.
The real infomax IC (Fig.\ \ref{fig:infomaxICA}) shows a clear increase in power near the median response time at about 300~ms, and a strong mean phase resetting which is visible near 300~ms as a phase-wrap (from $-\pi$ to $\pi$) and as a peak in the ITC.

The corresponding component obtained from real-map constrained-complex ICA at 5~Hz is displayed in Fig.\ \ref{fig:realspecICA}.
Its component map shares the spatial focus of maximum scalp projection with the time-domain IC map (cf.\ Fig.\ \ref{fig:infomaxICA}), but the spatial extent of the projection appears different.
Comparing the complex activation time-courses, the real-map constrained-complex IC shows a stronger response-locked power increase near 300~ms which is also more closely linked to the response time (Fig.\ \ref{fig:realspecICA}, center panel), and shows a more consistent phase-resetting and higher ITC near 300 ms after stimulus presentation (Fig.\ \ref{fig:realspecICA}, right panel).
This indicates that spectral-domain ICA may reflect subject behavior and underlying brain processes more faithfully than time-domain ICA.

The real part map obtained by decomposing the 5-Hz band with the fully-complex ICA algorithm (Fig.\ \ref{fig:cplxspecICA}, left) appears similar to the real-constrained component map.
The corresponding imaginary part map (Fig.\ \ref{fig:cplxspecICA}, second from left) has a non-negligible amplitude at the spatial focus of maximum scalp projection.
This indicates the presence of spatio-temporal dynamics in the data, and that these dynamics are modeled better with complex maps than with static real maps.
Here, the complex IC magnitude and phase activations (Fig.~\ref{fig:cplxspecICA}, right) do not appear qualitatively different from the activations obtained with the real-map constrained-complex algorithm (Fig.~\ref{fig:realspecICA}, right), although (as we have shown in section \ref{sec:quality-separation}) the fully-complex ICA results in IC activations with a higher degree of independence than those obtained with real-map constrained-complex ICA.

To illustrate the similarity of component maps over different spectral bands, Fig.\ \ref{fig:map_variability} displays those maps from the 10-Hz to 30-Hz decompositions that best match the illustrated 5-Hz component.
The maps in Fig.\ \ref{fig:map_variability} were obtained using the fully-complex ICA algorithm; only the magnitude maps are shown.
While the site of maximum scalp projection remains similar, the maps exhibit differences in shape and spatial extent, further suggesting that the complex spectral-domain ICA algorithm models aspects of the data that real ICA algorithms ignore.

Components not shown here are on the whole characterized by a single focus of activation in the associated scalp maps, which may indicate that their generators are located in spatially continuous (as opposed to disconnected) cortical regions.
About half of the components display a clearly non-zero imaginary part in their scalp maps, corresponding to processes with spatio-temporal dynamics.
However, we also find component maps with imaginary parts that do not appear to deviate significantly from zero.
Under the proposed model, these components correspond to static sources with negligible spatio-temporal dynamics.
This demonstrates that the complex ICA algorithm does not necessarily produce complex component maps, but also extracts the special case of real-valued mixing systems when supported by the data.

\section{Discussion and conclusion}
\label{sec:conclusion}

We have presented a new method for the analysis of dynamic brain data and in particular electroencephalographic signals.
The method is based on spectral decomposition of the sensor signals, and subsequent analysis within distinct spectral bands by means of a complex infomax algorithm for independent component analysis.

Although the applicability of ICA to time-domain EEG data is well established, the results obtained from the EEG dataset presented here---together with results from EEG data not shown---strongly support the applicability of complex spectral-domain ICA to EEG modeling and analysis.

Two different aspects of the method appear to offer improvements over previous ICA algorithms for modeling dynamic brain data.
First, signal superposition is modeled as a convolution, permitting sources to exhibit spatio-temporal dynamics.
Evidence for spatio-temporally dynamic patterns has been found in invasive recordings in animal cortex and includes spatial propagation of neural activity \citep{ArieliSGA_Science_1996}, traveling waves \citep{Freeman_MassActNervSys_1975,LopezS_chapter_1978}, and phase shifted activity between different regions \citep{stein00:_top}.
Second, signal superposition may be frequency dependent, allowing for distinct signal sources at different frequencies. 
This view follows naturally from the conventional notion of different frequency bands in EEG that appear to be related to different physiological functions.

The convolutive mixing assumption gained support from the decomposition results. 
The fully-complex spectral-domain ICA algorithm exhibited the lowest residual statistical dependencies, and many complex component maps showed clear non-negligible imaginary parts, indicating that complex ICA modeled spatio-temporal source dynamics in the data.
These first steps in understanding the relation  between complex independent components and underlying brain processes may be a qualitative step forward in modeling EEG data with ICA, a step that could potentially result in new insights into brain dynamics.

The assumption of frequency-dependent signal mixing was supported in three ways.
First, residual statistical dependencies after separation were lower with complex frequency-dependent ICA than with real wide-band ICA.
Second, component maps obtained with complex ICA varied across frequencies.
Third, results of complex ICA included distinct spectral ranges exhibiting clusters of similar independent components, which might pertain to physiological processes with activity over the corresponding spectral bands.
Compared to previous methods, our results indicate that improvements in analysis may be expected in the spectral range above 8~Hz, with largest improvements possible above 20~Hz, where the deviations between time-domain ICA and complex spectral-domain ICA results appear to be strongest.
However, the example data presented also indicated an advantage for complex ICA in the 5-Hz (theta) band, where complex ICA produced a component whose activity was more reliably related to subject task behavior than the corresponding real time-domain ICA component.

We have presented methods for assigning best-matching complex component pairs in different spectral bands to common sources.
For these data, complex ICA produced physiologically plausible component clusters.
However, further methodological improvements could be explored including other measures of component similarity, assignment procedures and quantitative clustering methods.

Other recording techniques, like the magnetoencephalogram (MEG) or functional magnetic resonance imaging (fMRI), and other electrical recordings from the human body such as electromyographic (EMG) and electrocardiographic (ECG) recordings, might also benefit from the presented methods.
Should statistical and physiological analysis of those data indicate the applicability of complex spectral-domain ICA, new directions for research might be open for those fields.

Several open questions regarding aspects of the presented methods should be investigated in further studies.

\begin{itemize}

\item 
The present work focused on the frequency-domain related aspects of the algorithm.
For a better understanding of the obtained components, it will be necessary to project them to the corresponding time-domain electrode voltages and study the resulting time-varying spatial distributions with respect to experimental design.
It should be possible to validate the proposed method by performing experiments for which a-priori knowledge exists as to expected spatio-temporal dynamics of the scalp maps.

\item 
One benefit of the frequency-dependent mixing assumption is that it may enable identification of a higher number of stable independent components.
However, each component obtained by the fully-complex algorithm can model only a single mode of spatio-temporal dynamics, corresponding to, e.g., a single direction of a cortical activation trajectory.
Sources with (nearly) identical foci of activation but different spatio-temporal dynamics may therefore be decomposed by complex ICA into distinct components.
Due to this higher sensitivity, the fully-complex algorithm could benefit from a larger number of EEG sensors.

\item
The stability of components both within and across subjects is a related area for further studies.
Variability of components with respect to, e.g., number of electrodes, recording-length, subject variability and variability across recording sessions should be investigated.

\item
The spectral basis employed in the present algorithms may appear as a natural choice, and it has the advantage of allowing for the simultaneous analysis of frequency-dependent and convolutive mixing using a single mathematical model.
However, other model choices may be possible and better adapted to the data than the present spectral basis.

\end{itemize}

The experimental results presented here indicate that complex spectral-domain independent components model aspects of spatio-temporal dynamics in the data that real-valued independent components ignore.
To support this possibility, we have shown one example showing spatio-temporal dynamics and a tighter relation of complex components to subject behavior.
To confirm the relevance of the new method for understanding brain data, it is important to further investigate the physiological plausibility of the decompositions and their functional relation to behavior based on more extensive analysis across subjects and experiments.

\addcontentsline{toc}{section}{References} 
\bibliographystyle{plainnat}
\bibliography{/home/jorn/doc/bib/bib,/home/jorn/doc/bib/eigene_pub}

\end{document}